\documentclass[12pt]{article}

\usepackage[left=1in,right=1in,top=1in,bottom=1in,letterpaper]{geometry}
\usepackage{sectsty}
\sectionfont{\Large}
\subsectionfont{\large}
\usepackage{caption,setspace}
\usepackage{booktabs}
\usepackage{wrapfig}
\usepackage{subcaption}
\usepackage{graphicx}
\usepackage{multirow,array}
\usepackage{arydshln}
\usepackage{algorithm}
\usepackage{algpseudocode}

\usepackage[authoryear]{natbib}

\usepackage{hyperref}
\hypersetup{colorlinks,urlcolor=blue,citecolor=black}
\usepackage{authblk}

\usepackage{booktabs}
\usepackage{wrapfig}
\usepackage{subcaption}
\usepackage{graphicx}
\usepackage{multirow,array}
\usepackage{arydshln}
\usepackage{algorithm}
\usepackage{algpseudocode}
\graphicspath{{Fig/}}

\usepackage{amsmath, amsthm, amssymb, amsfonts}

\numberwithin{equation}{section}
\usepackage[utf8]{inputenc}

\usepackage{comment}

\usepackage{amsmath, amsthm, amssymb, amsfonts}
\numberwithin{equation}{section}

\newcommand{\vone}[0]{\mathbf{1}}
\newcommand{\valpha}[0]{\boldsymbol\alpha}

\newcommand{\vxi}[0]{\boldsymbol\xi}
\newcommand{\vvartheta}[0]{\boldsymbol\vartheta}

\newcommand{\vSigma}[0]{\boldsymbol\Sigma}

\newcommand{\vzero}[0]{\mathbf{0}}
\newcommand{\vs}{\mathbf{s}}

\newcommand{\cT}{\mathcal{T}}

\newcommand{\tsin}{\textrm{sin}}
\newcommand{\tcos}{\textrm{cos}}

\newcommand{\CD}{\mathcal{D}}

\newcommand{\CF}{\mathcal{F}}
\newcommand{\CX}{\mathcal{X}}

\newcommand{\bdb}{{\pmb b}}

\newcommand{\bdx}{{\pmb x}}
\newcommand{\bdy}{{\pmb y}}

\newcommand{\bds}{{\pmb s}}

\newcommand{\BZ}{{\pmb Z}}
\newcommand{\BX}{{\pmb X}}

\newcommand{\BW}{{\pmb W}}

\newcommand{\bdalpha}{{\pmb \alpha}}
\newcommand{\bdbeta}{{\pmb \beta}}

\newcommand{\bdphi}{{\pmb \phi}}
\newcommand{\bdpsi}{{\pmb \psi}}

\newcommand{\bdnu}{{\pmb \nu}}

\newcommand{\bdomega}{{\pmb \omega}}

\newcommand{\E}{\mathrm{E}}

\newcommand{\MFM}{{\mathfrak M}}

\newcommand{\ben}{\begin{eqnarray}}
\newcommand{\een}{\end{eqnarray}}
\newcommand{\bse}{\begin{eqnarray*}}
\newcommand{\ese}{\end{eqnarray*}}
\newcommand{\bn}{\begin{enumerate}}
\newcommand{\en}{\end{enumerate}}

\newcommand{\black}[1]{{\color{black}#1}}

\newcommand{\wh}{\widehat}

\theoremstyle{thmstyleone}%
\theoremstyle{thmstyletwo}%
\theoremstyle{thmstylethree}%
\newtheorem{definition}{Definition}

\usepackage{xcolor}

\title{\Large{Spatially Varying Deep Functional Neural Network:\\
    Application in Large-Scale Crop Yield Prediction}}
\author[1]{Yeonjoo Park \thanks{Corresponding author: \texttt{yeonjoo.park@utsa.edu}}}
\author[2]{Bo Li}
\author[3]{Yehua Li}
\affil[1]{Department of Statistics and Data Science,\\
University of Texas at San Antonio}
\affil[2]{Department of Statistics and Data Science, Washington
University in St. Louis }
\affil[3]{Department of Statistics, University of California at Riverside  }
\date{}

\begin{document}

\maketitle

\begin{abstract}
Accurate prediction of crop yield is critical for supporting food security, agricultural planning, and economic decision-making. However, yield forecasting remains a significant challenge due to the complex and nonlinear relationships between weather variables and crop production, as well as spatial heterogeneity across agricultural regions. We propose a Spatially Varying Deep Functional Neural Network (SVD-funNet), a deep neural network architecture that integrates functional and scalar predictors with spatially varying coefficients and spatial random effects. The method is designed to flexibly model spatially indexed functional data, such as daily temperature curves, and their relationship to variability in the response, while accounting for spatial correlation.
SVD-funNet mitigates the curse of dimensionality through a low-rank structure inspired by the spatially varying functional index model (SVFIM). Through comprehensive simulations, we demonstrate that SVD-funNet outperforms state-of-the-art functional regression models for spatial data, when the functional predictors exhibit complex structure and their relationship with the response varies spatially in a potentially nonstationary manner.
Application to corn yield data from the U.S. Midwest demonstrates that SVD-funNet achieves superior predictive accuracy compared to both leading machine learning approaches and parametric statistical models. These results highlight the model's robustness and its potential applicability to other weather-sensitive crops.
\end{abstract}

\section{Introduction}

Corn is one of the most widely cultivated and consumed cereal crops worldwide, and it serves as a major agricultural commodity that underpins the livelihoods of farmers, drives agribusiness, and supports global markets. The United States, as the world’s leading producer and exporter of corn, depends heavily on the Midwest—often referred to as the U.S. Corn Belt—which accounts for the majority of national corn production. Accurate yield predictions are essential for balancing supply and demand, enabling farmers, investors, and policymakers to make informed decisions. Moreover, reliable forecasts play a key role in addressing global food security, given corn’s dual role as a dietary staple and a primary component of livestock feed.

Corn yield is highly sensitive to climate variability, as factors such as temperature and precipitation directly affect plant growth and productivity \citep{hatfield2011climate,Lobell2011,Huang2015}. \citet{Ray2015} estimated that climate variability explains approximately 60\% of corn yield variation in the American Midwest. Consequently, yield prediction often hinges on understanding the relationship between climate and crop growth \citep{Wong2019, Liu2022joe, Park2023}. However, this relationship is inherently complex and possibly heterogeneous across large geographic regions, making large-scale yield prediction a persistent challenge.

\begin{figure*}[!t]
\center
  \includegraphics[width=4in]{ 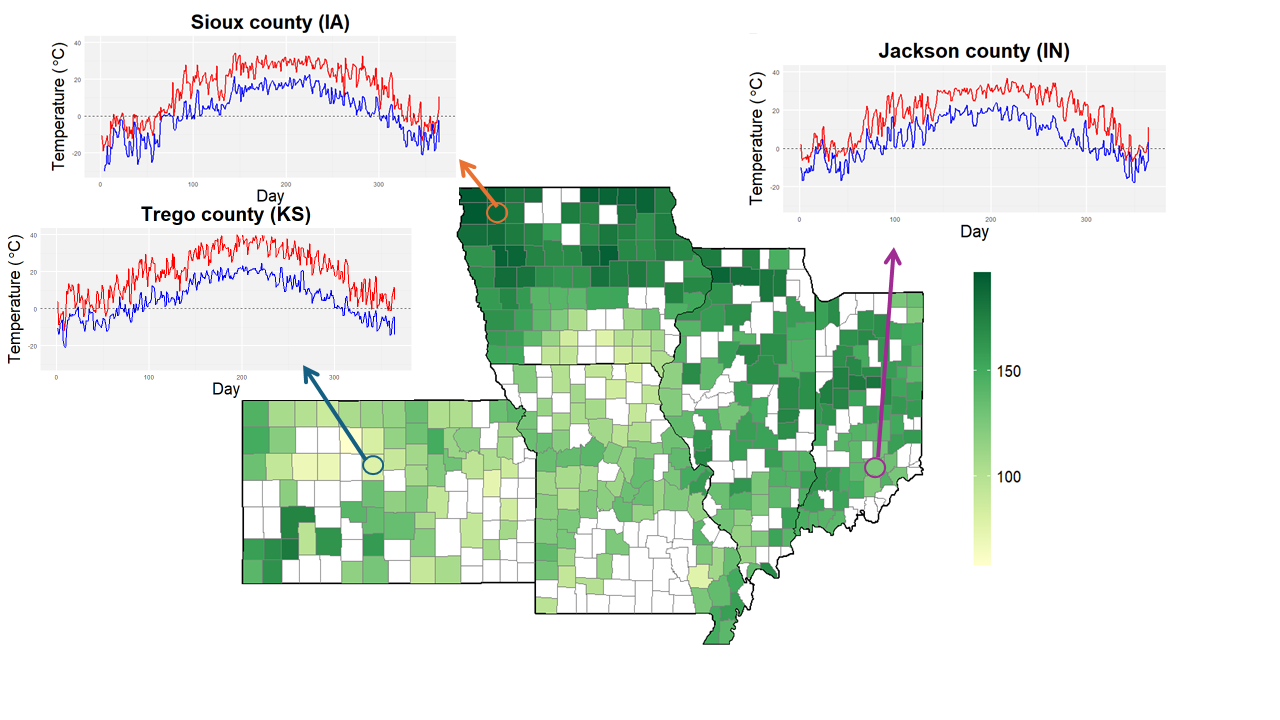}
   \caption{County level annual corn yield (measured in bushels per acre) in the Midwest region in 2010 and temperature trajectories in selected counties, where red and blue trajectories represent daily maximum and minimum temperatures (measured in $^\circ\mathrm{C}$), respectively. Counties shown in white indicate missing crop yield data.}\label{figure:intro_map}
\end{figure*}

Motivated by studying corn yield prediction, we collect county-level annual corn yield data (measured in bushels per acre) from 1999 to 2020 in the five Midwest states of Illinois, Indiana, Iowa, Kansas, and Missouri, through the National Agricultural Statistics Service (NASS) (\hyperlink{https://quickstats.nass.usda.gov/}{https://quickstats.nass.usda.gov/}). 
Due to a substantial number of missing values in the corn yield data after 2020 —likely resulting from disruptions caused by the COVID-19 pandemic — we exclude data beyond 2020 from our analysis.
Agricultural data are often unavailable in counties that are predominantly urban. Among the 102, 92, 99, 105, and 114 counties in these five states, we identify 403 counties with at least five years of recorded corn yield data during this period, including 79 in Illinois, 66 in Indiana, 93 in Iowa, 92 in Kansas, and 73 in Missouri.
We further obtain meteorological measurements between 1999 and 2020 for each county, including daily precipitation and daily maximum and minimum temperatures, from the National Climatic Data Center (NCDC) (\hyperlink{https://www.ncdc.noaa.gov/data-access}{https://www.ncdc.noaa.gov}). More details on the data can be found in \cite{Park2023}. Figure \ref{figure:intro_map} provides a graphical illustration of crop yield data across counties in the five Midwestern states. Additionally, we present sample trajectories of daily maximum and minimum temperatures from three randomly selected counties.

Since meteorological variables, such as maximum and minimum temperatures, influence crop growth on a daily basis, incorporating their yearly trajectories as functional predictors \citep{Ramsay2005} in crop yield prediction models can provide a more comprehensive representation of their impact.
For an overview of recent developments in functional data analysis (FDA), readers are referred to several review papers \citep{Morris2015, Wang-review2016, Li2022jmva}. For a theoretical foundation of FDA from the perspective of operator theory, we recommend \cite{Hsing-Eubank15}.
In the context of crop yield prediction, recent advances have employed both linear and nonlinear regression models that incorporate temperature curves as functional predictors \citep{Wong2019, Liu2022joe}. These models typically assume a homogeneous relationship between crop yield and temperature trajectories across all spatial locations. However, this assumption may be unrealistic for large geographical areas, as the effect of temperature on crop yield may vary regionally due to differences in local environmental conditions, such as soil moisture, soil pH, solar radiation, and wind velocity.
Since continuous monitoring of all potential confounding factors across a large spatial region is impractical, it is essential to account for spatial heterogeneity when using meteorological variables to predict crop yield. 

More recently, \cite{Park2023} proposed a spatially varying functional regression model (SVFM) that explicitly captures the spatially heterogeneous relationship between crop yield and weather data, demonstrating improved predictive performance. 
However, their method is fundamentally a parametric linear model, assuming that crop yield depends linearly on a few principal components of functional and multivariate predictors—albeit allowing the linear relationship to vary spatially as a stationary random process. Yet, several modeling assumptions may limit its predictive accuracy: whether the functional predictors can indeed be represented in a low-dimensional space, whether the linearity assumption is overly restrictive, and whether the assumption of stationarity appropriately captures the spatial structure. These limitations motivate us to develop a more robust while still powerful approach, particularly in light of growing evidence of nonlinear weather effects on crop yield \citep{Schlenker2006, schlenker2009nonlinear,Burke2015}. In this context, deep neural networks (DNNs) have emerged as a promising tool for making robust crop yield prediction \citep{kamilaris2018deep,Jabed2024}.

Recent theoretical work \citep{BauerKohler2019aos, schmidt2020nonparametric} has significantly advanced our understanding of DNNs as a nonparametric regression technique. For instance, \cite{BauerKohler2019aos} demonstrated that DNNs can circumvent the "curse of dimensionality" if the true underlying regression function has a low-rank structure, such as the generalized hierarchical interaction model. Furthermore, advances in customizing DNN architectures to improve model performance have been explored in various statistical applications \citep{sun2023individualized, ZhangLiSripadaKang2023jrssb}. DNNs with functional inputs have also been investigated by \cite{Thind2023, Rao2023, Wang2024}. However, none of these deep learning models explicitly account for spatial heterogeneity, a key challenge in large-scale crop yield prediction.

In this paper, we propose a Spatially Varying Deep Functional Neural Network (SVD-funNet) for predicting crop yield using both scalar and functional predictors (e.g., temperature trajectories). The proposed method leverages the power of deep neural networks to accommodate high-dimensional functional inputs, capture complex nonlinear relationships, and model flexible spatial dependency structures and heterogeneity. To construct our method, we first extend the SVFM of \cite{Park2023} to a class of Spatially-Varying Functional Interaction Models (SVFIM), which offer greater flexibility than traditional functional regression models commonly used in crop yield prediction. We then generalize SVFIM within the framework of generalized hierarchical interaction models, as studied in \cite{BauerKohler2019aos}. Drawing on techniques from \cite{Thind2023} to incorporate functional predictors, we show that the proposed SVD-funNet architecture adheres to the low-rank structure of SVFIM, thereby circumventing the curse of dimensionality and yielding strong predictive performance.
Following the idea of DeepKriging \citep{Chen2024}, SVD-funNet incorporates spatial random effects by embedding spatial basis functions as features within the neural network. Furthermore, by including interaction terms between spatial basis functions and both functional and scalar predictors, SVD-funNet enables spatially varying relationships between crop yield and the covariates.

The remainder of the paper is organized as follows.
Section \ref{sec:method} introduces SVD-funNet, which incorporates spatial basis functions to capture heterogeneous associations between inputs and responses, and accounts for spatial correlation via a spatial random effect. Section \ref{sec:sim} presents extensive simulation studies to evaluate the predictive performance of SVD-funNet in comparison with a functional regression model and an alternative deep learning approach. In Section \ref{sec:midwest_data}, we apply SVD-funNet to a comprehensive corn yield prediction study, benchmark it against various methods, and discuss insights gained from the prediction results. Finally, Section \ref{sec:conclusion} offers concluding remarks and directions for future research.

\section{Methodology}\label{sec:method}
To motivate the proposed Spatially-Varying Deep Functional Neural Network (SVD-funNet) architecture, we first review existing spatially varying functional regression models and discuss their potential extensions to a class of nonlinear Spatially-Varying Functional Interaction Models (SVFIM). We then describe how SVFIM can be implemented within a deep neural network (DNN) framework through the integration of spatial random effects and functional predictors.

\subsection{Spatially Varying Functional Interaction Model}\label{sec:svfim} 
For ease of exposition, we present the model based on data from a single year. Our analysis treats crop data from multiple years as conditionally independent replicates given the observed covariates, with rationales detailed in Section \ref{sec:midwest_data}. Let $Y(\bds)$ be the scalar response at location $\bds \in \CD$ for a spatial region $\CD \subset \mathbb{R}^2$, $\BX(\bds; t)=\{ X_1(\bds; t), \ldots, X_K(\bds; t)\}^\top$ defined for $t \in \cT$ denote $K$ functional predictors, and $\BZ(\bds)=\{ Z_1(\bds),$ $\ldots, $ $Z_J(\bds)\}^\top$ denote $J$ scalar predictors associated with $Y(\bds)$.  In our data, $Y(\bds)$ is the average corn yield per acre for the county located at $\bds$, $\CD$ is the spatial region of the five Midwestern states, and the time domain $\cT$ is a year. To predict $Y(\bds)$, \cite{Park2023} proposed the following Spatially Varying Functional Regression Model (SVFM), 
\begin{equation}\label{eq:svfm}
        Y(\bds) = \sum_{j=1}^J Z_j(\bds)\omega_j(\bds)  + \sum_{k=1}^K \int_\cT X_k (\bds;t) \beta_k(\bds;t) \textrm{d}t  + \eta(\bds) + \epsilon(\bds),  
\end{equation}
where $\bdomega(\bds)=\{\omega_1(\bds),$ $ \ldots\omega_J(\bds)\}^T$ is a vector of spatially varying coefficients for the scalar predictors, $\bdbeta(\bds;t) = \{\beta_1(\bds; t), \ldots, \beta_K(\bds; t)\}^T$ is a vector of spatially-varying functional coefficients, $\eta(\bds)$ represents spatial random effect, and $\epsilon(\bds)$ is white noise measurement error. When $\bdbeta(\bds; t) \equiv \bdbeta(t)$, $\bdomega(\bds)\equiv \bdomega$, and $\eta(\bds)\equiv 0$, model (\ref{eq:svfm}) reduces to the classic functional linear model, which has been intensively studied in the literature \citep{muller2005generalized,yao2005functional, li2007rates, Goldsmith2013, Reiss2017,zhou2023functional,Guo2025hdflm}.

{While the SVFM in (\ref{eq:svfm}) provides a flexible modeling framework, it ultimately assumes a linear relationship between crop yield and the functional predictors. It is well established that nonlinear models can yield substantial improvements in predictive performance. Moreover, prior work in functional regression has demonstrated that explicitly modeling interactions among projected functional components can further enhance predictive accuracy when such effects are present \citep{ChenMuller2012, UssetStaicuMaity2016, Liu2022joe}.
Together, these findings suggest that additional gains in crop yield prediction may be attainable through more flexible modeling strategies. This motivates us to consider nonlinear associations and interaction effects among covariates to enhance the SVFM. In the non-spatial regression setting, \citet{BauerKohler2019aos} proposed generalized hierarchical interaction models (GHIMs), which represent regression functions through a hierarchical composition of low-dimensional components based on projected scalar covariates and their interactions. Building on this idea, we extend the SVFM to a new class of Spatially Varying Functional Interaction Models (SVFIMs).
}

{
\begin{definition} ~
\begin{itemize}
    \item [(a)] A SVFIM of order $d^*$ and level 0 is written as
    \ben\label{eq:svfim_0}      
    \E \{Y(\bds) | \BX(\bds; \cdot), \BZ(\bds)\}= g^{[0]} \{ v_1^{[0]}(\bds), \ldots, v_{d^\ast}^{[0]}(\bds)\}, 
    \een
   where $v_i^{[0]}(\bds) = \int_\cT \BX^\top(\bds;t) \bdbeta_i(\bds;t) \textrm{d}t + \BZ^\top(\bds)\bdomega_i (\bds) + \eta_i(\bds)$, $i=1, \ldots, d^\ast$, and $g^{[0]}: \mathbb{R}^{d^*} \rightarrow \mathbb{R}$ is a smooth function. 
 \item [(b)] A SVFIM of order $d^*$ and level 1 is written as
    \ben\label{eq:svfim_1}      
    \E \{Y(\bds) | \BX(\bds; \cdot), \BZ(\bds)\}= \sum_{r=1}^{R_1} g_r^{[1]}\{ v_{1r}^{[1]}(\bds), \ldots, v_{d^\ast r}^{[1]}(\bds)\}, 
    \een
where $v_{jr}^{[1]}(\bds)=g_{jr}^{[0]}\{ v_{1jr}^{[0]}(\bds), \ldots, v_{d^\ast jr}^{[0]}(\bds)\}$, $j=1,\ldots, d^\ast$, are SVFIMs of order $d^*$ and level 0, $v_{ijr}^{[0]}(\bds) = \int_\cT \BX^\top(\bds;t) \bdbeta_{ijr}(\bds;t) \textrm{d}t + \BZ^\top(\bds)\bdomega_{ijr} (\bds) + \eta_{ijr}(\bds)$, $i=1, \ldots, d^\ast$, are the level 0 neurons in $v_{jr}^{[1]}(\bds)$, and $g_{r}^{[1]}, g_{jr}^{[0]}: \mathbb{R}^{d^*} \rightarrow \mathbb{R}$ are smooth functions. 
\item [(c)] A SVFIM of order $d^*$ and level $L$ is written as
\ben\label{defn_SVFIM} 
     \E \{Y(\bds) | \BX(\bds; \cdot), \BZ(\bds)\} = \sum_{r=1}^{R_L} g_r^{[L]}\{ v_{1r}^{[L]}(\bds), \ldots, v_{d^\ast r}^{[L]}(\bds)\},
\een
where $v_{jr}^{[L]}(\bds)$, $j =1,\ldots, d^\ast$ are SVFIMs of order $d^\ast$ and level $L-1$ defined recursively, with  $g_{r}^{[L]}: \mathbb{R}^{d^\ast} \rightarrow \mathbb{R} $ being smooth functions. 
  \end{itemize}
\end{definition}
These models are referred to as ``interaction models'' because they include general interactions between neurons at the same level. For example, $g_{ir}^{[\ell]}\{ v_{1}^{[\ell]}(\bds),$ $\ldots, v_{d^\ast}^{[\ell]}(\bds) \} = \sum_{j\neq j'} c_{jj'}^{[\ell]} \nu_{j}^{[\ell]}(\bds) \nu_{j'}^{[\ell]} (\bds)$ can be used to model two-way interactions between neurons in the $\ell$th level. The SVFIMs do not specify parametric forms for the interaction functions $g_{ir}^{[\ell]}$ and are therefore very flexible. The functional multiple-index model \citep{li2010deciding, chen2011single, radchenko2015index} is a special case of an SVFIM of level 0 under a non-spatial setting, where $v_i^{[0]}(\bds) \equiv v_i^{[0]}$ under $\bdbeta_i(\bds; t) \equiv \bdbeta_i(t)$, $\bdomega_i(\bds)\equiv \bdomega_i$, and $\eta_i(\bds)\equiv 0$. 
}
\subsection{Deep Neural Network}\label{sec:dnn}

{Deep neural networks (DNNs) are well known for their ability to approximate highly nonlinear regression functions. Among them, the multilayer perceptron (MLP), a fully connected feedforward DNN, is particularly recognized for its universal approximation capability for high-dimensional functions given sufficient network width. Notably, \citet{BauerKohler2019aos} show that MLPs are self-adaptive to a broad class of generalized hierarchical interaction models (GHIMs). This adaptivity allows MLPs to effectively approximate the SVFIM framework, which involves recursively defined structures with complex nonlinear dependencies and high-order feature interactions, without requiring prior knowledge of the underlying interaction structure and without suffering from the curse of dimensionality. 
}

{As a brief review of MLPs, consider using a MLP to approximate a high-dimensional nonparametric regression function, where the input vector is $\bdx \in \mathbb{R}^d$ and $d$ denotes the input dimension. For a network with $L$ hidden layers, neurons in the hidden layers are defined as 
\begin{eqnarray}\label{eq:mlp}
   \bdnu^{[\ell]} = \sigma_\ell(\BW_\ell \bdnu^{[\ell-1]} + \bdb_\ell), \quad \quad \ell =1, \ldots, L,
\end{eqnarray}
where $\bdnu^{[\ell]} \in \mathbb{R}^{n_\ell}$ is the output of the $\ell$-th layer, $\BW_\ell \in \mathbb{R}^{n_\ell \times n_{\ell-1}}$ is the weight matrix, $\bdb_\ell \in \mathbb{R}^{n_\ell}$ is the bias vector, and $\sigma_\ell(\cdot)$ is a component-wise activation function (e.g., ReLU, sigmoid, or tanh). Here, $\bdnu_0 = \bdx$ represents the input to the first hidden layer, and $n_\ell$ is the number of neurons in the $\ell$-th layer. From a structural perspective, the representation of $\ell$-th hidden layer $\boldsymbol{\nu}^{[\ell]}$ in \eqref{eq:mlp}, recursively defined in MLP, corresponds to the form of SVFIM of level $\ell$ defined with component functions $g_r^{[\ell]}(\cdot)$ introduced in Definition 1.

The final output layer produces the network's prediction
\ben\label{eq:mlp_out}
 y = \BW_{L+1}^\top \bdnu^{[L]} + b_{L+1},
\een
where $\BW_{L+1} \in \mathbb{R}^{n_L}$ and $b_{L+1} \in \mathbb{R}$ are the weight and bias of the output layer, respectively. Owing to the self-adaptivity of MLP to the GHIM structure, the width $n_\ell$ and depth $L$ of MLP do not have to match the order and level of the underlying GHIM. Such flexibility makes MLPs particularly appealing as a general methodology in high-dimensional nonparametric regression.}

The training of an MLP involves minimizing a loss function with respect to network parameters $\{\BW_\ell, \bdb_\ell\}_{\ell=1}^{L+1}$. This optimization is typically performed using gradient-based methods, such as stochastic gradient descent (SGD) or its variants, combined with backpropagation to compute the gradients efficiently.
In practice, the success of MLPs also depends on careful regularization and architectural choices to prevent overfitting and ensure generalization.

\begin{figure*}[!t]
\center
  \includegraphics[width=4.8in]{ 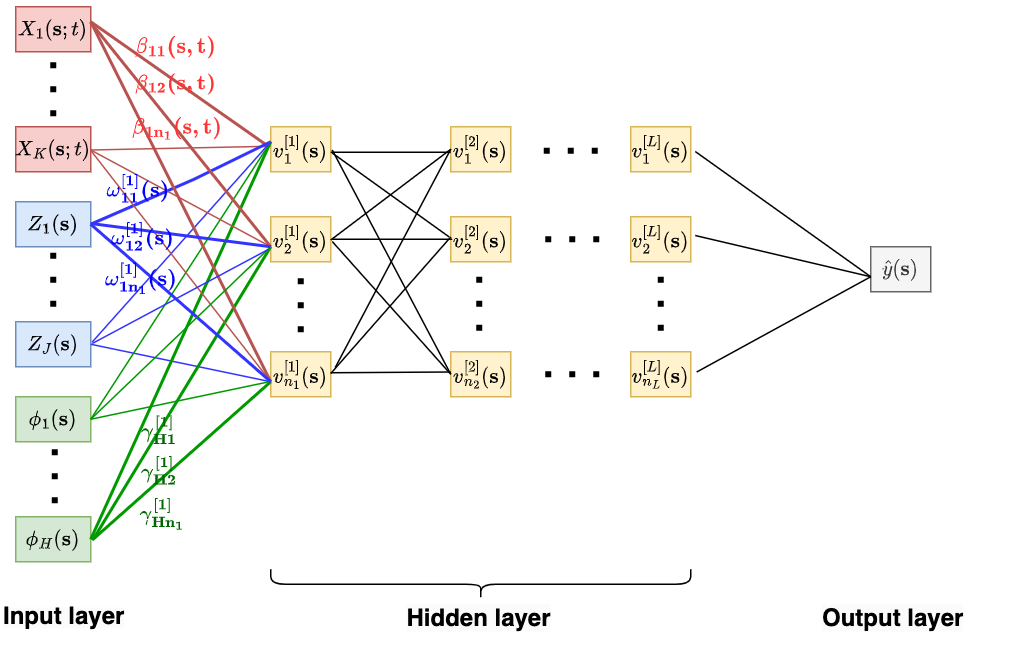}
   \caption{Structure of SVD-funNet with functional inputs, $X_k(\bds; t)$, scalar inputs, $Z_j(\bds)$, and layers characterizing spatial coordinates, $\phi_h(\bds)$. The location-specific weights for $X_1(\bds;t)$ and $Z_1(\bds)$ are highlighted in red and blue, respectively, with thick edges for the illustration. The green highlighted weights and thick edges illustrate spatial invariant weights on $\phi_H(\bds)$.}\label{figure:structure}
\end{figure*}
%
%
%
%

\subsection{Spatial Random Effect}\label{subsec:spatial_correction}
It is common for spatial prediction models such as (\ref{eq:svfm}) to include a spatial random effect $\eta(\bds)$ to account for unknown spatial variations in the response process, including those caused by unobserved confounders. In classic spatial statistics, $\eta(\bds)$ is typically modeled as a stationary, zero-mean Gaussian process with a parametric covariance structure \citep{Stein1999}. To accommodate large spatial datasets with complicated covariance structures, \cite{Nychka2015} proposed a multiresolution Gaussian process model based on which the spatial random effect can be modeled as $\eta(\bds) = \sum_{l=1}^L \delta_l(\bds)$, where  $\delta_l(\bds)$, $l=1, \ldots L$, denote $L$ independent spatial Gaussian processes. 
Each component is further modeled as $\delta_l(\bds) = \sum_{h=1}^{H_l} \omega_{h}^l \phi_{l,h}(\bds)$, where $\phi_{l,h}(\bds)$ are spatial basis functions and $\omega_{h}^l$ the corresponding coefficients.
Similar ideas were also used in the fixed-rank kriging proposed by \cite{Cressie2008}. 
\cite{Chen2024} incorporated these ideas into a deep kriging method to accommodate nonlinear prediction for nonstationary and non-Gaussian spatial data, where they fed spatial basis functions, $\bdphi(\bds)= (\phi_1, \ldots, \phi_H)^\top (\bds)$ as inputs in an embedding layer of deep neural network. 
To illustrate how the input of basis function represents spatial random effect in deep kriging, we can write $\eta(\bds)$ under a single hidden layer $(L=1)$ with $n_1$ neurons as, 
\begin{equation}\label{eq:deep_kriging}
  \eta(\bds) = \sum_{i=1}^{n_1} \sigma\bigg(\sum_{h=1}^H \omega_{ih} \phi_h(\bds)\bigg),
\end{equation}
which becomes a multiresolution Gaussian process model if the activation function $\sigma$ is linear. Compared with \cite{Nychka2015}, the deep kriging model (\ref{eq:deep_kriging}) can be considered as using the same set of basis functions for all latent Gaussian processes. \cite{Chen2024} shows that, by including flexible spatial basis functions and multiple layers, the deepkriging model provides very flexible modeling of a nonstationary spatial process $\eta(\bds)$.

Following the same rationale, we introduce the spatial random effect $\eta(\bds)$ to the proposed SVD-funNet architecture by including a set of spatial basis functions $\bdphi(\bds)$ as inputs in our neural network. Although there are many possible choices for $\bdphi(\bds)$, we adopt the multi-resolution thin plate spline (MRTS) basis functions advocated by \cite{Tzeng2018} for their ease of implementation. MRTS alleviates the challenges associated with basis function allocation, particularly when the data locations are irregular. Since MRTS basis functions are arranged in order of decreasing smoothness, from capturing global to local-scale features, they share similarities with Fourier basis functions and can effectively represent a smooth spatial function up to a specified resolution. \cite{Lin2023} empirically demonstrated that replacing Wendland functions \citep{Wendland1995} with MRTS basis functions enhances the spatial prediction performance of DeepKriging \citep{Chen2024}. Figure C.1 in the supplementary material illustrates the first 10 MRTS basis functions, which capture global variations, alongside the 41st to 50th MRTS basis functions, which capture local variations, based on 40 equally spaced inner knots selected from the spatial domain in the real data application. 
\subsection{Deep Spatial Neural Net with Functional Input}\label{sec:dsnn}

Inspired by recent work on functional deep learning \citep{Thind2023} and deep kriging \citep{Chen2024}, we propose a Spatially Varying Deep Functional Neural Network (SVD-funNet), which takes functional inputs and can be applied to estimate all functional regression models discussed in Section \ref{sec:svfim}.
For a neural network with $L$ hidden layers and $n_l$ neurons at each level, given inputs located at $\bds$ including 
functional covariates, $X_1(\bds;t), \ldots, X_K(\bds;t)$, for $t\in \cT$ and scalar covariates, $Z_1(\bds), \ldots, Z_J(\bds)$, neurons in the first layer of the proposed SVD-funNet are
\begin{equation}
 v_i^{[1]}(\bds) 
 = \sigma\Bigg( \sum_{k=1}^K \int_\mathcal{T}\beta_{ik}(\bds; t)  X_{k}(\bds;t)  \mathrm{d}t + \sum_{j=1}^J \omega_{ij} (\bds)Z_j(\bds) \nonumber \\
  + \eta_i(\bds) + b_i\Bigg),
 \label{eq:fnn_first_layer} 
\end{equation}
for $i=1,\ldots, n_1$. Here, $\beta_{ik}(\bds;t)$ and $\omega_{ij} (\bds)$ are interpreted as location-specific weights for the functional and scalar predictors, respectively, $\eta_i(\bds)$ is the spatial random effect, and $b_i$ is the intercept or bias term. The remaining $L-1$ hidden layers and the output layer are defined in the conventional way as in (\ref{eq:mlp}) and (\ref{eq:mlp_out}). The architecture of the proposed DSNN is illustrated in Figure \ref{figure:structure}.

In the first hidden layer $v_i^{[1]}(\bds)$, we approximate the unknown functions in the right hand side of (\ref{eq:fnn_first_layer}) using basis functions, similar to \cite{Thind2023}. 
Specifically, we write the location-specific coefficient function $\beta_{ik}(\bds; t)$ as
\begin{equation}\label{eqn:beta_basis_expansion_time}
    \beta_{ik}(\bds; t) = \sum_{m=1}^{M_k} c_{ikm}(\bds) f_{km}(t),   
\end{equation}
where $\{f_{km}(t)\}_{m=1}^{M_k}$ is a set of basis functions to represent the $k$-th functional weight. These basis functions can be fixed basis functions such as splines, Fourier basis functions, or wavelets \citep{Ramsay2005, Thind2023} or data-driven bases such as the empirical principal components of $\BX$ \citep{Liu2022joe,Park2023}. 
We further assume that location-specific loadings $\{c_{ikm}(\bds) \}$ are smooth functions of $\bds$ which can be written as $c_{ikm}(\bds) = \sum_{p=1}^{P} \kappa_{ikmp}\psi_{p}(\bds)$, where $\bdpsi(\bds)=(\psi_1, \ldots, \psi_P)^\top(\bds)$ is a set of spatial basis functions defined over $\bds \in \CD$. With this representation, (\ref{eqn:beta_basis_expansion_time}) can be written as
\begin{equation}\label{eqn:beta_basis_expansion_space_time}
    \beta_{ik}(\bds; t) = \sum_{m=1}^{M_k} \sum_{p=1}^{P} \kappa_{ikmp} \psi_{p}(\bds) f_{km}(t).   
\end{equation}
Similarly, we express the spatially-varying weight function $\omega_{ij}(\bds)$ using the same set of basis functions
\begin{equation}\label{eqn:omega_basis_expansion}
    \omega_{ij}(\bds) = \sum_{p=1}^{P} \vartheta_{ijp}  \psi_p(\bds).  
\end{equation}

{Note that $\bdpsi(\bds)$ is used to model spatially-varying effects of the scalar and functional predictors and can be different from the basis function $\bdphi(\bds)$ used for spatial random effects. Both $P$ and $H$, dimensionalities of $\bdpsi(\bds)$ and $\bdphi(\bds)$, respectively, are determined through the hyperparameter selection strategies detailed in Section \ref{subsec:model_spec}.} Our empirical studies show that the prediction performance of the proposed SVD-funNet is not sensitive to the choice of spatial basis functions as long as the resolution numbers $P$ and $H$ are sufficient to capture the spatially-varying effects.

Despite the high dimensionality inherent in SVD-funNet, we show that the curse of dimensionality can be alleviated when a low-rank structure, such as that in the SVFIM, is present. Further elaboration is provided in Section A of the supplementary material.

\subsection{Model Specification and Parameter Tuning}
\label{subsec:model_spec}
Together, the form of the $i$-th neuron in the first hidden layer of Figure \ref{figure:structure} is written as
\begin{equation} 
         v_i^{[1]}(\bds) 
 =  \sigma  \Bigg( \sum_{k=1}^K \sum_{m=1}^{M_k} \sum_{p=1}^P \kappa_{ikmp} \psi_p(\bds) \int_\mathcal{T}f_{km}(t) X_{k}(\bds;t)\mathrm{d}t + \nonumber\\
  \sum_{j=1}^J \sum_{p=1}^P \vartheta_{ijp}  \psi_p(\bds) Z_j(\bds) + \sum_{h=1}^H \gamma_{ih} \phi_h(\bds)  + b_i \Bigg),
\end{equation}\label{eqn:neuron_all}
where the integral in \eqref{eqn:neuron_all} can be approximated with a numerical integration method for each of the $K$ functional inputs. As mentioned in Section \ref{subsec:spatial_correction}, the rest of the $L-1$ hidden layers of the network follow the form of the conventional neural network model. To fit this network, we employ a backpropagation algorithm using the Adam Optimizer \citep{Kingma2014} for the implementation. Section 2.2 of \cite{Thind2023} provides a sketch of a general optimization scheme for stochastic gradient descent on the model with functional input. As pointed out by \cite{Thind2023}, representing functional covariates and functional weights by basis functions not only respects the continuity of the functional covariates but also increases model efficiency compared to models using discrete observations on the functional covariates as multivariate inputs.

Hyperparameter tuning is critical for optimizing neural network performance, encompassing standard parameters such as the number of layers, neurons per layer, activation functions, learning rate, decay rate, validation split, epochs, batch size, and early stopping criteria. In our SVD-funNet, we further introduce architecture-specific hyperparameters including: the choice of basis functions (e.g., Fourier or Splines) defined over $t$; the expansion size $M$ for functional weights (assumed consistent across all $K$ functional inputs); and the dimensionalities $P$ and $H$ of spatial MRTS basis functions, where $P$ controls spatial variability in functional and scalar weights at the first hidden layer while $H$ determines spatial random process approximation. For parameter optimization, we employ $C$-fold cross-validation, where we iteratively train on $C-1$ folds and evaluate on the remaining fold to compute the mean square prediction error (MSPE). This process repeats $
C$ times to ensure robust performance assessment across all data partitions.

\section{Simulation studies}
\label{sec:sim}
We conduct simulation experiments under various scenarios to evaluate the predictive performance of our method in comparison to other cutting-edge functional regression and functional deep learning methods.

\subsection{Data Generation}\label{subsec:sim_data_gen}
We adopt the 403 counties across five states from our real dataset as the spatial domain in the simulation and generate data using the following model:  
\black{
\begin{equation}\label{eqn:sim_model}
Y_d(\bds_l)= g \Big\{Z_d(\bds_l) \alpha(\bds_l) + \int_{\mathcal{T}} X_d(\bds_l;t) \beta(\bds_l;t) \textrm{d}t + \eta(\bds_l) \Big\} + e_d(\bds_l),
\end{equation}
}

where $l=1, \ldots, n_d$ are indices of counties and $d=1, \ldots, 5$ are replicate years. The spatial random effect $\eta(\bds)$ represents county-to-county variations that do not change over the years. When $g(\cdot)$ is the identity function, model (\ref{eqn:sim_model}) reduces exactly to the data generation model used in \cite{Park2023}. Moreover, model (\ref{eqn:sim_model}) is a special case of SVFIM in (\ref{defn_SVFIM}). Below, we provide details on the data generation process for each component of (\ref{eqn:sim_model}). 

\vskip 2mm
\noindent - {$X_d$, $\beta$, $Z_d$ and $\alpha$:} 
We consider two scenarios for the functional and scalar covariates. 
\begin{itemize}
    \item {\it Scenario 1 (Low-rank feature on functional covariates with stationary spatial dependence).} The first scenario generates spatially dependent functional covariates under a low-dimensional representation,
$
X_d(\bds_l;t) = \sum_{r=1}^4 \xi_{dr}(\bds_l) f_r(t),
$  
$t \in \mathcal{T}=[0,1]$, using four basis functions $f_r(t)$ and their corresponding loadings $\xi_{dr}(\bds_l)$. 
The spatially varying coefficient is further generated from $\beta(\bds;t) = \sum_{r=1}^4 \vartheta_{r}(\bds_l) f_r(t)$ using the same $f_r(t)$. 
This reduces the integral part in (\ref{eqn:sim_model}) into a low dimensional structure: $\sum_{r=1}^{4}\xi_{dr}(\bds_l){\vartheta}_r(\bds_l)$. 
This structure aligns with the underlying model assumptions for SVFM in \cite{Park2023}. 

The basis functions are set as $f_1(t) = \sqrt{2}\tsin(2\pi t)$, $f_2(t)=\sqrt{2}\tcos (2\pi t)$, $f_3(t)=\sqrt{2}\tsin(4\pi t)$, and $f_4(t)=\sqrt{2}\tcos(4\pi t)$. We choose to have $n_d \equiv n=403$ for all $d$, and generate the spatially correlated loadings $\xi_{dr}(\bds_l)$ from Gaussian processes such that 

$\vxi_{dr}$ $=$ $\{\xi_{dr}(\bds_1),$ $\ldots,$ $ \xi_{dr}(\bds_{n})\}^T \sim N(\vzero_n, \lambda_r \vSigma(\zeta_r) \}  $, $r=1,\ldots, 4$, where $(\lambda_1, \ldots, \lambda_4)= (4,2,1,0.5)$. Here, $\vSigma(\zeta_r)$ are correlation matrices governed by Mat\'ern correlation functions $\rho_r(h)= \{\Gamma(\tau) 2^{\tau-1}\}^{-1} (h / \zeta_r)^{\tau} K_{\tau}(h/ \zeta_r),$ where $h$ is the distance between two counties,  $\tau$ is the smoothness parameter, $\zeta_r$ is the range parameter, and $K_\tau(\cdot)$ is the modified Bessel function of the second kind \citep{Stein1999}. Specifically, we set $\tau=1$, $\zeta_1=400$, $\zeta_2=300$, $\zeta_3=200$, and $\zeta_4=100$.
Given that distances between Midwest counties range from 16 to 1530 km with an average of 516 km, the values of our range parameters represent relatively moderate correlation structures, matching the parameters estimated by \cite{Park2023}. 

To generate $\beta(\bds_l;t)$, we simulate its coefficients as Gaussian processes such that $\vvartheta_1\sim N\{ 2\cdot \vone_n, \vSigma(\zeta_1)\}$,  $\vvartheta_2\sim N\{-2\cdot \vone_n, \vSigma(\zeta_2)\}$,  $\vvartheta_3 \sim N\{ \vone_n, \vSigma(\zeta_3)\}$, and $\vvartheta_4\sim N\{- \vone_n, \vSigma(\zeta_4)\}$, where $\vvartheta_r = \{\vartheta_r(\bds_1), \ldots,\vartheta_r(\bds_n)\}^T$. For the scalar covariate, we generate $Z_d(\bds_l) \overset{iid}{\sim}\text{unif}(0,2)$ and the coefficient $\alpha(\bds_l)$ by a Gaussian process $\valpha = \{ \alpha(\bds_1), \ldots, \alpha(\bds_l)\}^T $ $ \sim N\{\vone_n, \vSigma(\zeta_1)\}$. 

\item {\it Scenario 2 (Real data covariates).} The second scenario considers a more realistic setting by directly borrowing maximum temperature trajectories from our real data for $X_d(\bds_l;t)$, so that its underlying structure may not be represented by just a few basis functions, and possibly exhibits a more complicated spatial dependency among trajectories. Specifically, we use the data from 2005 to 2009, which has relatively low proportions of missing counties. While Scenario 1 considers the same number of spatial locations at each $k$, this scenario has varying $n_d$ ranging from 315 to 345, and a total of 380 locations are used in modeling. We set the scalar covariate $Z_d(\bds_l)$ as the annual precipitation from the real data. The functional and scalar coefficients, $\beta(\bds_l;t)$ and $\alpha(\bds_l)$, are generated using the same procedure described in Scenario 1. 
\end{itemize}

\noindent - $\eta(\cdot)$: 
We generate spatial random effects via $\eta(\bds) = \sum_{h=1}^{10} \upsilon_{h} \phi_h(\bds)$, where $\phi_h(\bds)$ are orthonormal multi-resolution thin plate spline (MRTS) basis functions, and $\upsilon_{h} \sim N(0, 1)$. This setting ensures that the contribution of $\eta(\bds)$ to the variability of response is comparable to that of scalar and functional covariates, so that no single component dominates the variability of responses. Figure C.1 in the supplementary material illustrates 10 MRTS basis functions used in the experiment. 

\vskip 2mm
\noindent - $g(\cdot)$: 
We consider the following four link functions in our simulation study.
\begin{itemize}
    \item[1.] Linear function: $g(x) = x$.
    \item[2.] Double exponential function: $g(x) = c_e \exp(-|x|/2)$.
    \item[3.] Sine function: $g(x) = c_s \sin(x)$.
    \item[4.] Piecewise linear function: 
    \[g(x) = 
\begin{cases}
c_{p_1} \{ 1+(x+c_{p_2})/ c_{p_3} \}, & \text{if } x < -c_{p_2} ;\\
c_{p_1}, & \text{if }   |x| \leq c_{p_2};\\ 
c_{p_1} \{ 1 - (x - c_{p_2})/ c_{p_3} \}, & \text{if } x > c_{p_2}.
\end{cases}
\]
\end{itemize}
Among the four choices, the sine function exhibits the most striking nonlinear behavior. The piecewise linear function is perhaps the most realistic for crop yield prediction, as it can capture saturation effects. For instance, precipitation generally benefits crop growth, but excessive rainfall can lead to flooding, which may damage or destroy crops.
The constants $c_e$, $c_s$, $c_{p_1}$, $c_{p_2}$, and $c_{p_3}$ are set to ensure sufficient nonlinearity for the given $x$ values as well as similar variation of $g(\cdot)$ for each scenario. Under Scenario 1, we set $c_e=10$, $c_s=3$, $c_{p_1}=6$, $c_{p_2}=2$, and $c_{p_3}=3$. For Scenario 2, we set $c_e=9$, $c_s=3$, $c_{p_1}=7$, $c_{p_2}=1$, and $c_{p_3}=5$. 
These choices yield a variance of approximately 5 for $g(\cdot)$, under each scenario.  

\vskip 2mm
\noindent - $e_d$: 
The random errors are generated by $e_d(\bds_l) \overset{iid}{\sim} N(0, \sigma_e^2)$ with  $\sigma_e^2$ set either at 3.33 or 2 to approximate the signal-to-noise ratio (SNR) at 1.5 and 2.5, respectively. SNR is defined as the ratio between the variance of $E(Y|X, Z, \eta)$, around 5 in our experiment, and $\sigma_e^2$. These choices roughly align with the early findings that 60\% of the variability in corn yield in the American Midwest can be explained by climate variability \citep{Ray2015}.

\subsection{Implementation and Evaluation Metrics}\label{subsec:sim_implement_comparison}

 \begin{table}[b!]
  \begin{center}
  \caption{The averages of mean squared prediction errors (MSPE) with standard errors in parentheses, computed over 100 replications. Boldface indicates the best performance (the lowest MSPE) for each combination of model scenario, choice of $g(\cdot)$, and signal-to-noise ratio (SNR) level.}\label{tab:sim-res-rev}
  \resizebox{\textwidth}{!}{
\begin{tabular}{cccccccccc}
\toprule
& &  \multicolumn{2}{c}{{SVD-funNet} (proposed)}  &  \multicolumn{2}{c}{SVFM}  & \multicolumn{2}{c}{FNN} & \multicolumn{2}{c}{\black{DKF}} \\ 
\cmidrule(lr){3-4} \cmidrule(lr){5-6} \cmidrule(lr){7-8} \cmidrule(lr){9-10}
 & $g(\cdot)$ & SNR=1.5 &  SNR=2.5 & SNR=1.5  & SNR=2.5 &  SNR=1.5 &  SNR=2.5 &  SNR=1.5 &  SNR=2.5 \\
\midrule
\multirow{4}{*}{\shortstack{Scenario 1}} & Linear  & 4.76 (0.08)  & 3.23 (0.07)  &  {\bf 3.15 } (0.05) & {\bf 3.14 } (0.05) & 6.39 (0.10) & 4.92(0.11)    & 5.21 (0.09)   & 3.52 (0.08)\\
&Piecewise linear  &  5.03 (0.10) & 4.32 (0.09)  & {\bf 3.57} (0.12) & {\bf 3.43 } (0.09)  & 7.12 (0.17) & 6.59 (0.16) & 6.70 (0.18)  & 4.94 (0.17) \\
&Double exponential   & 5.40 (0.07) & 3.93 (0.05)  &{\bf 3.68 } (0.07) & {\bf  3.57 } (0.06)  & 7.04 (0.10) & 5.62 (0.09) &  6.16 (0.08) & 4.41 (0.07) \\
&Sine   & 5.93 (0.06) & 4.21 (0.05) & {\bf 3.90 } (0.06) & {\bf 3.70 }(0.06) & 7.59 (0.06) & 6.21 (0.05) &  7.39 (0.06) & 5.87 (0.09) \\
\midrule
\multirow{4}{*}{\shortstack{Scenario 2}} &Linear  & {\bf 4.72 } (0.05)  &  {\bf 3.01 } (0.03)  & 5.37 (0.07) & 3.88 (0.06) & 6.39 (0.15) & 4.98 (0.14)  & 5.28 (0.06)  & 3.39 (0.04)\\
&Piecewise linear  &   {\bf 4.72 } (0.06) & {\bf 3.02} (0.04)  & 5.52 (0.10) & 3.94 (0.09)  & 6.28 (0.18) & 5.88 (0.19) &  5.23 (0.07) & 3.39 (0.04) \\
&Double exponential   &{\bf 5.02 } (0.09) &  {\bf 3.57 } (0.08)  & 6.19 (0.13) & 4.56 (0.11)  & 7.48 (0.17) & 6.06 (0.16)&  6.03 (0.14)   & 4.19 (0.10) \\
&Sine   &{\bf 5.98 } (0.08) &  {\bf 4.55 } (0.06) & 7.01 (0.09) & 5.40 (0.08) & 7.19 (0.07) & 5.80 (0.06)  & 6.83 (0.08)  & 5.06 (0.08) \\
\bottomrule
\end{tabular}
}
  \end{center}
\end{table}    

The simulation is repeated 100 times at each combination of model scenario and choice of $g(\cdot)$, under two levels of SNR. In each run, 20\% of the observations are randomly selected as test data, with the remaining 80\% used for training. 
We choose Fourier basis functions for $f_{dm}(t)$ in (\ref{eqn:neuron_all}).  Hyperparameters of our SVD-funNet model include
the number of basis functions for each expansion, the dimensionalities of MRTS basis functions used to capture spatial variation in weight parameters and to model spatial random effects, the number of hidden layers, and the number of neurons per hidden layer. Since our simulated data mimics the real data, we adopt the same hyperparameter selection procedure for the real data, which will be described in detail in Section \ref{subsec:data_model}.

\black{We compare the predictive performance of our SVD-funNet with three cutting-edge methods: the spatially varying functional regression model (SVFM) \citep{Park2023}, the functional neural network (FNN) method \citep{Thind2023}, and DeepKriging with functional input (DKF).} The SVFM has demonstrated superior performance in predicting corn yield compared to other functional regression models by incorporating spatially varying functional and scalar coefficients. The implementation of SVFM requires selecting the dimension parameter $p$, the number of functional principal components (FPCs) to be included in the model. In Scenario 1, we use the true dimension $p = 4$, which reflects prediction under the known true dimensionality and represents the optimal performance achievable by SVFM. For Scenario 2, we set $p = 5$, following the optimal dimensionality identified by \cite{Park2023} using the same dataset.
The FNN is a state-of-the-art neural network approach designed to incorporate functional inputs, although it does not account for spatial dependencies. To train FNN, we use a hyperparameter selection strategy similar to that of SVD-funNet to determine the optimal configuration at each simulation iteration. \black{The DKF is an extension of DeepKriging \citep{Chen2024} that incorporates functional inputs. It can be viewed as an intermediate model between FNN and the proposed SVD-funNet, but it is not equipped to incorporate spatially varying model parameters. The DKF is also tuned using the same hyperparameter selection approach.}

We evaluate the prediction performance using the Mean Squared Prediction Error (MSPE), calculated as 
\[
 \sum_{(k,l) \in \mathcal{A}_{\textrm{test}}} \{ Y_d(\bds_l) - \hat{Y}_d(\bds_l)\} ^2/  ~|\mathcal{A}_{\textrm{test}}|,
\]
where $\mathcal{A}_{\textrm{test}} = \{(k,l): Y_d(\bds_l)$ belongs to the test set\}, and $|\mathcal{A}_{\textrm{test}}|$ is its cardinality. 
By comparing the predictive performance of SVFM and the proposed SVD-funNet across various settings, we gain insight into the conditions under which deep learning models outperform flexible parametric methods. Furthermore, the comparison between SVD-funNet and FNN enables us to empirically assess the advantages of incorporating spatial information into deep learning algorithms.

\begin{figure*}[!t]
\centering
    \includegraphics[width=5in]{ 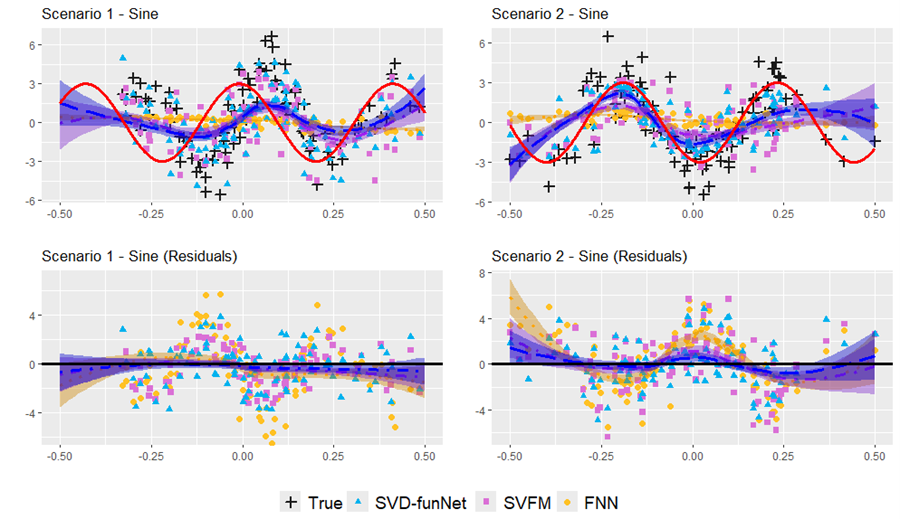}
  \caption{ Top panel: Estimated $g$ function from SVD-funNet (blue), SVFM (purple), and FNN (yellow) under Scenarios 1 and 2, based on a randomly selected simulation run under SNR = 2.5. To aid visualization, 100 data points from the test set are randomly selected for display. The $x$-axis represents ${Z_d(\bds_l) \alpha(\bds_l) + \int_{\mathcal{T}} X_d(\bds_l;t) \beta(\bds_l;t) \textrm{d}t + \eta(\bds_l)}$ from \eqref{eqn:sim_model}, and the $y$-axis represents $\hat{Y}_d(\bds_l)$. For clarity, the data points are centered and scaled, and local regression curves (long dashed lines) with corresponding standard error bands are superimposed. The solid red line indicates the true sine function for $g(\cdot)$, and the black "+" symbols denote the true values of $Y_d(\bds_l)$.
  Bottom panel: Corresponding residual plots for the top panel, also superimposed with local regression curves. The solid black line represents the zero line on the y-axis. 
  } \label{figure:sim_link}
\end{figure*}
\subsection{Simulation Results}
\label{subsec:sim_result}

Table \ref{tab:sim-res-rev} displays the average MSPE together with its standard error over 100 simulation runs, evaluated on the test set. 
To illustrates the recovery of the function $g(\cdot)$ by each method, 
Figure \ref{figure:sim_link} uses the sine function as an example to compare the performance of each method under the two scenarios. The upper panel plots ${Z_d(\bds_l) \alpha(\bds_l) + \int_{\mathcal{T}} X_d(\bds_l;t) \beta(\bds_l;t) \textrm{d}t + \eta(\bds_l)}$ on the x-axis and $\hat{Y}_d(\bds_l)$ on the y-axis, based on a subset of the test set from a randomly selected run. The bottom panel displays the corresponding residuals. 
To better visualize the differences among the three methods, we first applied local regression to the points for each method and then superimposed the estimated regression means along with their corresponding standard errors on each plot. The results for double exponential and piecewise linear are deferred to Figure B.1 of the supplementary material.

Under Scenario 1, SVFM consistently outperforms the other three models, regardless of the nonlinearity in
$g(\cdot)$ or the level of SNR. This result is somewhat surprising, as we initially expected that SVFM, being a linear model, would struggle to capture the nonlinear relationships introduced by a nonlinear $g(\cdot)$. However, upon closer reflection, this outcome can be explained by the flexibility of SVFM’s spatially varying coefficients, which allow it to effectively capture both linear and nonlinear patterns — provided that the variability occurs across spatial locations. However, if the nonlinearity is localized within individual locations rather than varying spatially, SVFM will not be able to model such relationships.
This experiment reveals that when the response is related to the functional covariates through a low-dimensional feature and the spatial correlation in the data is stationary, as assumed in SVFM, a flexible parametric model like SVFM can be highly effective.

The left column of Figure \ref{figure:sim_link} confirms that SVFM predictions align closely with the true $g(\cdot)$. The SVD-funNet predictions also follow the trend of $g(\cdot)$ well, though with slightly reduced accuracy near the right end. Additionally, the residuals from SVD-funNet appear to exhibit slightly greater variance compared to those from SVFM. Nevertheless, we note that our experimental setup places the SVFM model in a favorable position, as the model fitting directly uses the true value $p = 4$. As such, the performance of SVFM may be somewhat overly optimistic.

More importantly, the overly simplified structure assumed in Scenario 1 is unlikely to hold in most real-world applications. Under Scenario 2, where the functional covariates are drawn directly from real observations that likely cannot be adequately represented using only a few basis functions and may exhibit more complex spatial dependency structures, the predictive accuracy of SVD-funNet uniformly surpasses that of SVFM. This suggests that SVD-funNet is better equipped to extract meaningful information from functional covariates with complex structures than the parametric SVFM. The right column of Figure \ref{figure:sim_link} shows that SVD-funNet captures the trend of $g(\cdot)$ better than SVFM. The residuals from SVD-funNet are more centered around zero and exhibit slightly less variability compared to those from SVFM.

The FNN, which does not account for spatial correlation or spatial variability, consistently underperforms compared to both SVD-funNet and SVFM. As shown in Figure \ref{figure:sim_link}, predictions from FNN tend to cluster around the mean of the responses, regardless of the form of the $g(\cdot)$ function. This behavior arises because FNN, by not accounting for the spatial characteristics of data, tends to average information across all locations, resulting in fitted values that gravitate toward the global mean. \black{ Although DKF outperforms FNN by incorporating evaluated spatial basis functions alongside functional covariates in the input layer, resulting in smaller MSPEs, it still underperforms the proposed SVD-funNet, which further incorporates spatially varying parameters.}

In terms of computation, the average time to train the proposed SVD-funNet model and make predictions under its optimally selected configuration is approximately 1.15 minutes per simulation run for Scenario 2, while SVFM is fitted through Bayesian hierarchical modeling and requires an average of 81.31 minutes to generate 500 MCMC samples. This represents a substantial difference in computational burden.
 
In sum, the flexible SVFM can capture nonlinear relationships between inputs and responses when the nonlinearity can be absorbed into the spatial variability of those relationships. SVFM performs well under correct model specification, that is, when the functional predictors can be represented in a low-dimensional space and their relationship with the response varies spatially according to a stationary process.
In such settings, SVD-funNet may underperform relative to SVFM, potentially due to overfitting caused by overparameterization.
However, real-world data rarely conform to such simplified structures. For more complex and realistic datasets, SVD-funNet clearly outperforms SVFM in predictive accuracy and offers a substantial advantage in computational efficiency compared to the Bayesian implementation of SVFM. While FNN handles functional data effectively, it lacks the capacity to model spatial correlation and heterogeneity, limiting its performance for spatially structured data.

\section{Midwest crop yield prediction} \label{sec:midwest_data}
We apply the proposed SVD-funNet to model and predict county-level average corn yield (in bushels per acre) across five Midwestern states - Illinois, Indiana, Iowa, Kansas, and Missouri - during the period 1999–2020. In practice, year-to-year variability in crop yields is largely influenced by a combination of factors, including seed genetics, biotechnology, and management decisions. Additionally, agricultural practices such as land fallowing and crop rotation further contribute to annual variation.
To account for these effects, we first remove the year-specific component from the corn yield by subtracting the annual average, resulting in yield anomalies. We then model these annual anomalies as conditionally independent (in time) random processes given the meteorological covariates. This approach is consistent with the practice in \cite{Park2023}, which empirically demonstrated a lack of temporal correlation in demeaned yield data across years.

In our modeling framework, we use county-level daily maximum and minimum temperatures as functional covariates and monthly averages of precipitation as scalar covariates. Although raw daily precipitation records were available, we explored several strategies for incorporating this data, including annual, quarterly, and monthly averages, as well as treating it as a functional covariate. Among these options, using monthly average precipitation yielded the best predictive performance across both our model and the comparison models. We attribute this to the inherent characteristics of daily precipitation data, which is often zero-inflated with occasional sharp spikes, unlike typical functional data that exhibit smooth and continuous patterns. Moreover, using monthly averages is a practical choice, as crop yield is more strongly influenced by the cumulative effect of water availability over time than by short-term fluctuations.

\begin{figure*}[t!]
\center
  \includegraphics[width=4.8in]{ 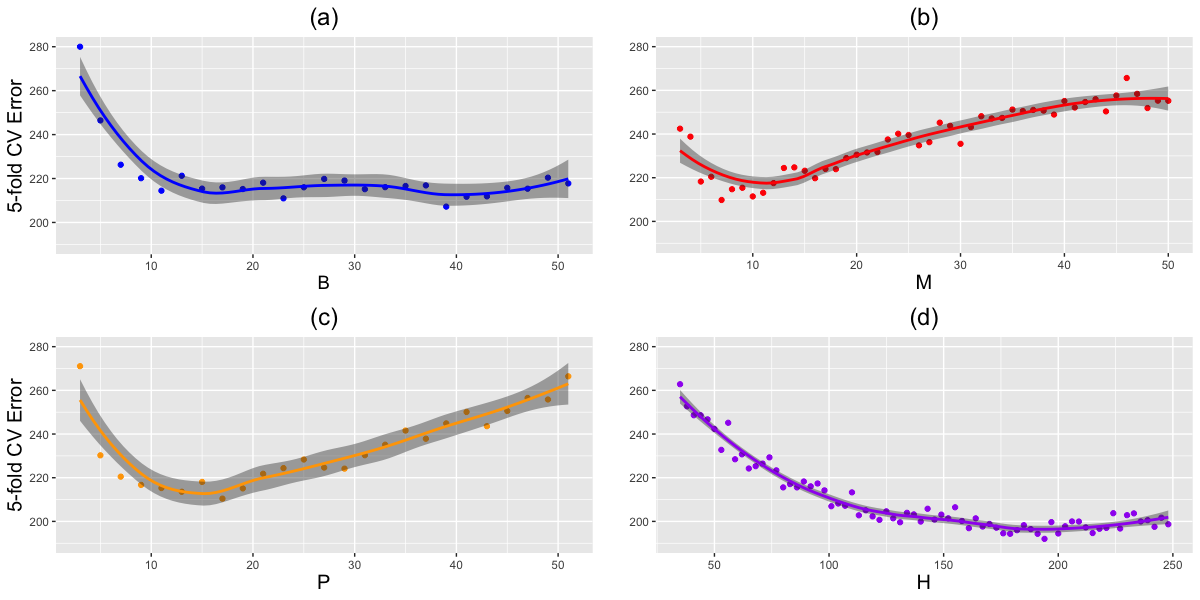}
   \caption{ Sensitivity analysis of key hyperparameters based on 5-fold cross-validation (CV) errors for 
(a) number of Fourier basis functions $B$ to represent the functional input;
(b) number of Fourier basis functions $M$ for modeling temporal variation in the functional weights;
(c) number of MRTS basis functions $P$ for modeling spatial variation in the functional weights; and
(d) number of MRTS basis functions $H$ for modeling the spatial random effect process.
Each panel is overlaid with a local regression curve, with shaded regions indicating the associated standard errors.
}\label{figure:tune}
\end{figure*}
\subsection{SVD-funNet Model Training} \label{subsec:data_model}

We first apply standard preprocessing to the maximum and minimum temperature curves by registering them on $B$ Fourier basis functions, with the goals of maintaining as many functional features as possible while denoising \citep{Ramsay2005}. We represent the location-specific functional weights $\beta_{ik}(\bds; t)$ in \eqref{eqn:beta_basis_expansion_space_time} using Fourier basis functions $f_{km}(t)$. In this study, we adopt a shared set of Fourier basis functions across $k$, i.e., $f_{km}(\cdot) = f_m(\cdot)$. Then, MRTS basis functions are employed for $\psi_p(\bds)$ in \eqref{eqn:beta_basis_expansion_space_time} and \eqref{eqn:omega_basis_expansion} to model spatially varying coefficients for both functional and scalar weights, respectively. Additionally, MRTS basis functions are also used for $\phi(\bds)$ in \eqref{eqn:neuron_all} to model spatial random effects. All dimensions of basis functions are treated as hyperparameters.

Standard deep learning hyperparameters, such as learning rate, decay rate, number of epochs, and validation split, have been found to exert minimal influence on prediction performance when set within reasonable ranges. Consequently, we fix these at standard values, and also adopt the popular sigmoid activation function, in line with \citep{Thind2023}. We identified several key hyperparameters that significantly affect prediction accuracy. These include: the number of Fourier basis functions ($M$) used to capture temporal variability in $\beta_{ik}(\bds,t)$ within the first hidden layer; the number of MRTS basis functions ($P$) for modeling spatial variability in $\beta_{ik}(\bds,t)$; the number of basis functions ($H$) for approximating the spatial random effect process $\eta_i(\bds)$; the number of hidden layers ($L$); and the number of neurons per layer ($N$). We determine the optimal configuration of these hyperparameters through cross-validation.

To illustrate the sensitivity of prediction accuracy to the number of basis functions, $B$, $M$, $P$, and $H$, we conducted experiments to examine how the corresponding 5-fold cross-validation (CV) errors vary with each of these parameters. Although the number of Fourier basis functions $B$ is not a hyperparameter of the SVD-funNet model, it can still affect prediction accuracy and is therefore included in this analysis. \black{To isolate the effect of each hyperparameter, we vary one parameter at a time while fixing all others at values suggested by earlier experiments. For hyperparameters that have not yet been selected, we manually explore their effects over a predefined set of candidate values. Although different combinations of the remaining hyperparameters may inflate or deflate the overall 5-fold cross-validation errors associated with a given hyperparameter, we found that the general pattern, particularly the pronounced error dips over certain ranges, remains consistent.} The results are presented in Figure~\ref{figure:tune}. For $B$, $M$, and $P$, the range from 3 to 51 sufficiently captures the trend in CV errors as the parameter increases. In contrast, for $H$, we use a wider range (35 to 250) since the CV errors decrease more gradually with increasing $H$.

Figure~\ref{figure:tune}(a) shows that the effect of $B$, the number of Fourier basis functions used to expand $X_k(\bds, t)$, on prediction accuracy stabilizes after an elbow point at approximately $B = 15$. This suggests that, once $B$ is large enough to capture the dynamics of the functional predictor, the model’s performance becomes relatively insensitive to its precise value. Based on these exploratory results, we fix $B = 21$ for registering the functional inputs in model training and no longer treat it as a hyperparameter.
In Figures~\ref{figure:tune}(b) and (c), the CV error initially decreases with increasing $M$ or $P$ and then rises after reaching a minimum. This indicates that using too many basis functions in the functional weights, whether for temporal variability using Fourier basis ($M$)  or for spatial variability using MRTS ($P$), can lead to overfitting and degrade predictive performance. These observations highlight the importance of tuning $M$ and $P$ during model training.
Figure~\ref{figure:tune}(d) shows that CV error remains relatively stable once a sufficiently large value of $H$ is reached, although a slight increase in error is observed for very large values of $H$, suggesting possibly mild overfitting.

\begin{table*}
\small
\setlength{\tabcolsep}{1pt}
\centering
        \caption{MSPE and weighted MSPE from 10-fold cross-validation using SVD-funNet, variations of the Functional Neural Network (FNN), Neural Network (NN), eXtreme Gradient Boosting (XGB), and the Spatially Varying Functional Regression Model (SVFM). A ``$\checkmark$" indicates inclusion of the corresponding component in the model, while an ``x" denotes its absence. The labels (i), (ii), and (iii) denote the types of enhancements applied to each model. 
        }
\begin{tabular}{@{}l*{7}{c}}
   \toprule  
    & \multicolumn{3}{c}{Temperature covariates} & \multicolumn{2}{c}{Spatial terms} \\
    \cmidrule(lr){2-4}\cmidrule(lr){5-6}
     \multicolumn{1}{p{1cm}}{\centering Learning \\ Model } & \multicolumn{1}{p{1.5cm}}{\centering Function} & \multicolumn{1}{p{1.8cm}}{\centering Multi- \\variate} & \multicolumn{1}{p{1.8cm}}{\centering FPC\\ score} & \multicolumn{1}{p{1.8cm}}{\centering Spatially varying \\ weights }& 
     \multicolumn{1}{p{1.9cm}}{\centering Spatial \\ random effects } & \multirow{2}{1cm}{MSPE} & \multicolumn{1}{p{1cm}}{\centering Weighted\\~~~ MSPE }\\
       \midrule
         \midrule

          \begin{tabular}{l} SVD-funNet\\
                (Proposed)
                \end{tabular} & $\checkmark$&  x &  x & $\checkmark$  & $\checkmark$ &{\bf 186.4} & {\bf 133.4}\\
       \midrule
        \multirow{3}{*}{\begin{tabular}{l} FNN\\
             FNN(i)\\
             FNN(ii)
                \end{tabular}} &$\checkmark$ & x  & x & x  &  x & 484.6 & 367.3\\ 
         & $\checkmark$ & x & x & x  & $\checkmark$ & 283.8 & 192.3\\ 
         & $\checkmark$& x & x & $\checkmark$   & x & 291.5& 192.5\\ 
       \midrule
        \multirow{4}{*}{\begin{tabular}{l} NN\\
             NN(i)\\
             NN(ii)\\
             NN(iii)
                \end{tabular}} &x & $\checkmark$ & x & x  & x  & 455.5& 335.9\\ 
         & x &x  & $\checkmark$ & x  &x & 539.2& 415.4 \\ 
         & x&x  &$\checkmark$ &  $\checkmark$ & x& 286.4& 197.9\\   
         & x&x  & $\checkmark$&  $\checkmark$ & $\checkmark$& 258.2& 189.5\\    
       \midrule
     \multirow{4}{*}{\begin{tabular}{l} XGB\\
                XGB(i)\\
                XGB(ii)\\
                XGB(iii)
                \end{tabular}}  &x & $\checkmark$ & x & x  &  x & 409.8& 306.8\\ 
         & x &x  & $\checkmark$ & x  & x & 479.8& 369.3\\ 
         & x&x  &$\checkmark$ &  $\checkmark$ & x& 253.9& 184.5\\   
         & x&x  & $\checkmark$& $\checkmark$ & $\checkmark$& 202.4& 157.0\\   
       \midrule
  \begin{tabular}{l} SVFM\\
                \end{tabular} & $\checkmark$&  x &  x & $\checkmark$  & $\checkmark$ &425.5 &  355.7\\
          \bottomrule
    \end{tabular}
    \label{tab:crop_predict}
    \end{table*}

Figure~\ref{figure:tune} not only demonstrates the individual effects of each of the four hyperparameters but also helps identify ranges for $M$, $P$, and $H$ to guide parameter optimization. In the subsequent analysis, we determine the optimal combination of hyperparameters each time SVD-funNet is trained by performing a grid search within narrowed ranges informed by Figure~\ref{figure:tune}, thereby reducing computational cost. Specifically, the grid search is conducted over the following sets: $M \in \{5, 7, 9, 11, 13\}$, $P \in \{5, 7, 10, 12, 15, 20\}$, $H \in \{100, 130, 150, 180, 200, 250\}$, $L \in \{4, 5, 6, 7, 8\}$, and $N \in \{16, 32, 64\}$.

\subsection{Models for Comparison}
To comprehensively evaluate the proposed SVD-funNet method and gain deeper insight into its strengths, we also apply four additional approaches to the data: the Functional Neural Network (FNN) and Spatially Varying Functional Regression Model (SVFM) discussed in Section \ref{sec:sim}, along with two widely used machine learning techniques, Neural Network (NN) \citep{lecun2015deep} and eXtreme Gradient Boosting (XGB) \citep{chen2016xgboost}.

FNN serves as a submodel of SVD-funNet that excludes spatial variability and spatial correlation. To isolate the contributions of these spatial components, we also examine two intermediate submodels of SVD-funNet: (i) FNN with spatially varying parameters and (ii) FNN with spatial random effects. Combining (i) and (ii) yields the full SVD-funNet model. By comparing these submodels to both FNN and SVD-funNet, we aim to quantify the individual contributions of spatially varying parameters and spatial random effects to the model's ability to capture variability in the data.

Neither NN nor XGB was originally designed to incorporate functional inputs or account for spatial structure in the data. To adapt these models to functional covariates, we begin by representing the raw temperature curves as multivariate inputs, specifically, 730-dimensional vectors corresponding to daily maximum and minimum temperatures over a year. To make these methods more comparable to SVD-funNet, we progressively enhance their baseline formulations through three tiers of modification. 
(i) Functional dimension reduction: Rather than using raw multivariate inputs, we first reduce the dimensionality of the functional covariates via functional principal component (FPC) scores. We retain 21 FPC scores that explain over 98\% of the total variability in the temperature trajectories. This results in a special case of the FNN model, where both the functional inputs and their associated weights are represented using the same set of FPC basis functions. In contrast, our SVD-funNet model provides greater flexibility by allowing distinct and more general basis functions for representing the inputs and the weights.
(ii) Spatially varying parameters: We enhance the input layer by introducing interaction terms between the FPC scores and the MRTS basis functions, thereby enabling NN and XGB to model spatially varying parameters.
(iii) Spatial random effects: Finally, we add spatial random effects by including the spatial basis functions $\phi_1(\bds), \ldots, \phi_H(\bds)$ in the input layer.
The NN model enhanced through all three tiers, i.e., NN(iii), can be viewed as a special case of SVD-funNet, where both the functional inputs and functional weights are represented using FPC basis functions. By comparing this enhanced NN to SVD-funNet, we can evaluate the additional benefits of allowing for more general and flexible choices of basis functions in modeling functional data and their associated weights.

For a fair comparison, we use the same set of MRTS basis functions of dimension $(P)$ and the same number of spatial basis functions ($H$) as in SVD-funNet for all enhancements applied to FNN, NN, and XGB. For completeness, we also include SVFM as a representative of parametric statistical model, given its strong predictive performance in Section \ref{sec:sim}, especially when the underlying data structure lies in a low-dimensional space. 
Note that we run the SVFM  with maximum and minimum functional temperature covariates and an annual average of precipitation as the scalar covariates, rather than the monthly average, due to extreme computation time with the increase in model dimensionality.

\subsection{Prediction Assessment and Implication} \label{subsec:data_result}
\begin{figure*}[t!]
  \centering
  \includegraphics[width=5in]
  { 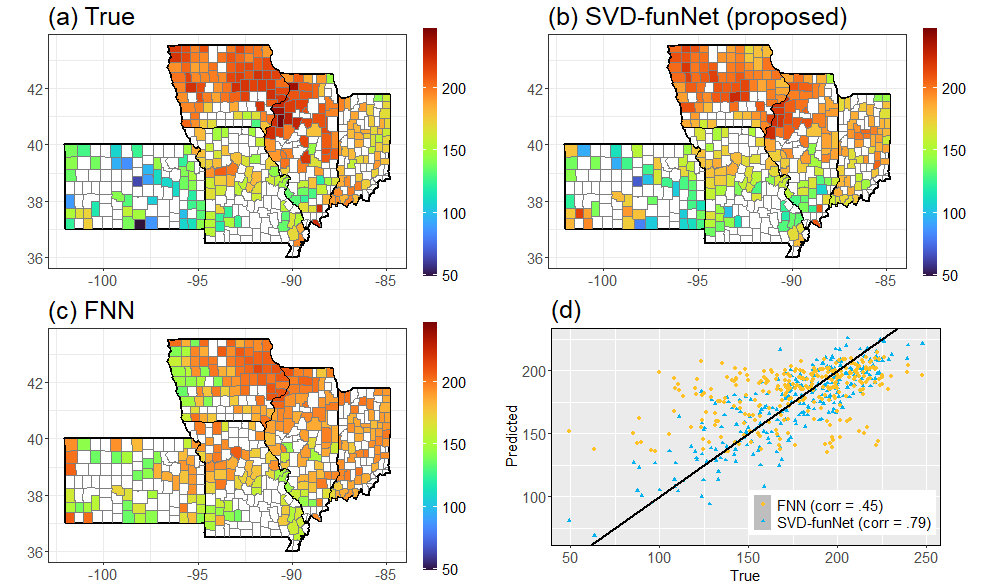}
  \caption{(a) The county-level crop yield (bushels per acre) collected in 2018 and predicted corn yields using 1999-2017 data from (b) the proposed SVD-funNet method and (c) FNN, respectively, where the blank represents counties with missing data; (d) A scatter plot between true and predicted corn yields from each method.}\label{figure:prediction}
\end{figure*}

We evaluate the predictive performance of the proposed and competing models using 10-fold cross-validation, in which all year–county combinations are randomly divided into 10 equal subsets. Each fold takes turn to serves as the test set once, while the remaining nine folds are used for training. For each test set, we train the models on the corresponding training data and compute the prediction errors on the test data. For SVD-funNet, the optimal combination of hyperparameters is selected via grid search over the candidate sets specified in Section \ref{subsec:data_model}.
Two types of MSPE are calculated for model comparison: (i) Regular MSPE as defined in the simulation study, and (ii) Weighted MSPE using the size of the harvest land as the weight, calculated by averaging
    $$\sum_{(k,l) \in \mathcal{A}_c}\pi_k(\bds_l)\{Y_k(\bds_l) - \hat Y_k(\bds_l)\}^2/\sum_{(k,l) \in \mathcal{A}_c}\pi_{k}(\vs_l),$$
over $c=1,\ldots,10$, where $\mathcal{A}_c=\{(k,l); ~Y_k(\bds_l) \mbox{ belongs to the}$ $c$th test set\} and $\pi_k(\bds_l)$ denotes the size of harvested land (acre) in year $k$ and county $l$. The harvest land size information is also obtained from the National Agricultural Statistics Agency. 
We consider weighted MSPE as it may reflect the practical importance of accurate predictions in counties with larger agricultural output.  Table~\ref{tab:crop_predict} reports both regular and weighted MSPEs for all models under comparison.

The proposed SVD-funNet significantly outperforms all competing methods, achieving the lowest MSPE and weighted MSPE. The basic FNN model, which lacks any spatial adaptation, yields the highest errors in both metrics. \black{Two additional enhanced variants—FNN(i), which incorporates spatial random effects, and FNN(ii), which incorporates spatially varying weights—are also considered for comparison. We note that FNN(i) is identical to DKF, which is included in the simulation study; however, we retain the FNN(i) label here to maintain a systematic comparison among FNN and its variants FNN(i)–(ii).} Both enhanced FNN variants show substantial improvements in predictive accuracy, highlighting that each spatial component independently contributes to explaining variability in corn yield. As expected, combining both components, as in the full SVD-funNet model, results in even greater predictive performance than including either one alone

However, its two enhanced variants - FNN(i), incorporating spatial random effects, and FNN(ii), incorporating spatially varying weights  - show substantial improvements in predictive accuracy. These results highlight that each spatial component independently contributes substantially to explaining variability in corn yield. As expected, combining both components, as in the full SVD-funNet model, yields even greater predictive performance than including either one alone.

\begin{figure*}[t!]
  \centering
\includegraphics[width=4.8in]
  { 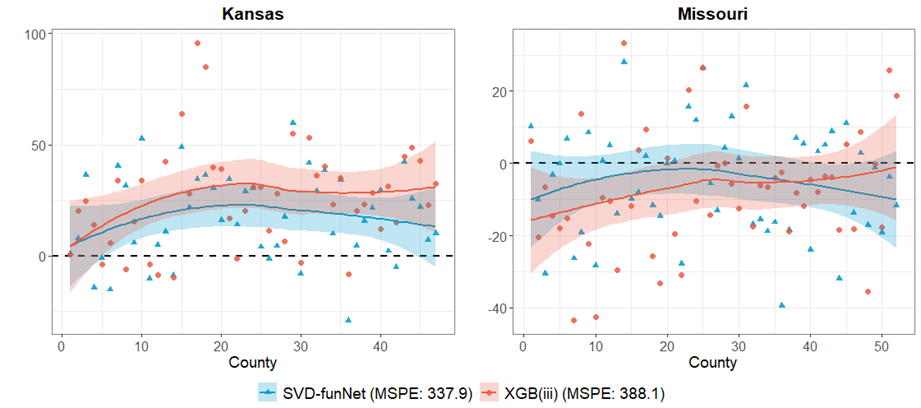}
  \caption{\black{Prediction residuals across counties in Kansas and Missouri from the model to predict 2018 corn yields using 1999-2017 data from the proposed SVD-funNet (blue) and the XGB(iii) method enhanced with spatially varying weights on FPC scores and spatial random effects (red). The corresponding local regression curves are superimposed, with shaded regions indicating their associated standard errors.}}
  \label{figure:prediction2}
\end{figure*}

The other machine learning methods, NN, XGB, and their variants, exhibit a similar pattern: prediction accuracy improves when spatially varying weights are assigned to the FPC scores, and further improves when spatial random effects are incorporated. However, neither NN(iii) nor XGB(iii), the versions that include both spatial components, outperform the proposed SVD-funNet. This comparison highlights the advantage of SVD-funNet’s flexible functional data modeling, which does not rely on FPC functions and allows the use of distinct basis functions for representing functional inputs and their associated weights.
Interestingly, in the absence of spatial modeling, both NN and XGB achieve lower cross-validation (CV) errors when using raw daily maximum and minimum temperature curves as multivariate inputs, compared to using multivariate FPC scores.
While incorporating FPC scores into DNNs has proven effective for some applications \citep{wang2023deep}, this approach does not yield comparable benefits for our data. We conjecture that for our data, the FPC scores derived solely based on the variance structure of the functional covariates without considering their relationship to the response variable, may not be able to adequately capture the variation in the response \citep{jolliffe1982note}. 

Finally, the SVFM shows notably inferior predictive performance compared to the deep learning approaches that incorporate spatial structure, exhibiting substantially higher MSPEs and considerably longer computation times. Relative to its performance in the simulation study, even under Scenario 2, the SVFM performs markedly worse on real data. We think this is likely because the true structure of spatially indexed temperature trajectories and the relationship between both functional and scalar covariates and corn yield is likely more complex than what a parametric model, even a flexible one like SVFM, can capture. Even when real covariates are used in Scenario 2, the simulation model for the response may still be overly simplistic and biased in favor of SVFM. For instance, the weights on the functional and scalar covariates in the real setting may not follow a stationary Gaussian process. These findings highlight the importance of leveraging the flexibility and robustness of deep learning methods when dealing with complex, high-dimensional climate-agriculture data. While using monthly average precipitation as scalar covariates, rather than annual precipitation as in the current model, might improve predictive performance, we do not anticipate it would reduce the MSPE by half. Rather, it would significantly increase the computational burden.

To visually illustrate the difference in predictive performance, we use data from 1999 to 2017 to predict crop production in 2018. Although data are available through 2020, we choose 2018 for prediction to facilitate clearer visualization, as the years 2019 and 2020 contain a relatively high proportion of missing yield data. We first show the difference between the proposed SVD-funNet and the basic FNN (without spatial enhancements) in Figure \ref{figure:prediction}.
SVD-funNet predictions in panel (b) clearly capture the spatial structure observed in the true responses shown in panel (a). While the basic FNN recovers some spatial patterns, its predictions in panel (c) perform poorly for both high and low yield regions. The scatter plots in panel (d) further highlight the advantage of SVD-funNet: its predictions align closely with the true observations, achieving a correlation coefficient of 0.79. In contrast, the FNN model exhibits substantial deviation from the observed values, with a much lower correlation coefficient of 0.45. The scatter plot also reveals that FNN predictions tend to cluster around the mean of the observations. This phenomena  arises from FNN’s inability to account for spatial heterogeneity, resulting in a fitted model that tend to average
the covariate-response relationship across all counties, rather than capturing localized behaviors.

We then compare SVD-funNet with the best-performing machine learning model, XGB(iii), as identified in Table \ref{tab:crop_predict}. 
Overall, the MSPEs for SVD-funNet and XGB(iii) are 337.9 and 388.1, respectively, and their weighted MSPEs are 286.7 and 291.1. Among the five states, SVD-funNet outperforms XGB(iii) in Illinois (IL), Kansas (KS), and Missouri (MO), yields comparable results in Indiana (IN), and only underperforms in Iowa (IA).  Detailed results are provided in Table C.1 of the supplementary material.
\black{
To illustrate differences in predictive performance, Figure~\ref{figure:prediction2} presents county-level prediction residuals for Kansas (KS) and Missouri (MO), overlaid with local regression curves and shaded bands representing their standard errors. The results show that residual trends from SVD-funNet remain closer to zero across counties in both states and exhibit less variability than those from XGB(iii), suggesting that SVD-funNet yields predictions with reduced bias and improved precision. Corresponding results for the remaining three states are provided in Figure C.2 of the Supplementary Material.}

\section{Concluding remarks} \label{sec:conclusion}

We propose SVD-funNet, a deep neural network designed for spatial prediction with functional and scalar covariates. Its architecture incorporates spatial basis functions to model spatially varying functional and scalar network parameters, as well as spatial random effects.
Although SVD-funNet appears to be high-dimensional, we show that the curse of dimensionality is mitigated when the underlying structure conforms to a low-rank SVFIM representation.
Beyond theoretical justification, SVD-funNet achieves substantial improvements in prediction accuracy for large-scale corn yield forecasting across the U.S. Midwest. It outperforms the state-of-the-art parametric functional regression model with spatially varying coefficients by \cite{Park2023}, as well as other deep neural networks, including those augmented to capture spatial structure and accommodate functional covariates. While developed for corn yield prediction, SVD-funNet is broadly applicable to other crops whose growth is strongly influenced by weather patterns, such as soybeans \citep{schwalbert2020satellite}. 

Our extensive simulation results also offer valuable insights into the relative strengths of parametric statistical models and deep learning approaches. While SVD-funNet performs exceptionally well with spatially indexed functional covariates characterized by complex structures and spatial dependencies, the SVFM proves effective when the data reside in a low-dimensional space with a stationary spatial structure. The spatially varying coefficients in SVFM allow it to capture nonlinear relationships between the response and covariates, provided that the nonlinearity manifests across spatial locations. These findings underscore the complementary strengths of statistical and deep learning models, and they highlight the need for more comprehensive studies comparing various approaches to spatial prediction with functional covariates, akin to the benchmarking efforts undertaken for time series forecasting \citep{makridakis2018statistical}.

\black{Lastly, we note that recently proposed Spatial Bayesian Neural Networks (SBNNs; \citealp{Zammit2024}) share conceptual similarities with our approach through the use of a spatial embedding layer, although their primary focus is on modeling a spatial process $Y(\mathbf{s})$ without covariates via a Wasserstein loss. Extending SBNNs to incorporate scalar and functional covariates would yield a Bayesian counterpart of our SVD-funNet, enabling formal statistical inference and uncertainty quantification, and represents a promising direction for future research.
Conformal prediction provides another attractive framework for uncertainty quantification. Recent advances have been developed for time series and spatiotemporal data \citep{mao2024valid, jiang2024spatial}. However, conformal methods specifically tailored to our setting of prediction for spatially aggregated data have not yet been established, highlighting an important avenue for future work. More broadly, uncertainty quantification for deep learning models remains a critical and challenging area of ongoing research.}

\bibliographystyle{chicago} 
\bibliography{ref-bib}       

\newpage
\def\theequation{S\arabic{section}.\arabic{equation}}
\def\thesection{S\arabic{section}}
\def\thetable{S\arabic{table}}
\def\thefigure{S\arabic{figure}}

\clearpage
\setcounter{page}{1}  
\setcounter{section}{0}

\begin{center}
{ \Large \center {\bf {Supplementary Material for \\``{\color{black} Spatially Varying Deep Functional Neural Network: \\
Application in Large-Scale Crop Yield Prediction}''}}}\\
\end {center}

The Supplementary Material provides theoretical justification of the SVD-funNet architecture under a low-rank structure from a class of SVFIM. It also contains additional results from simulation studies and the crop yield prediction application.

\section{Theoretical Justification of the Proposed Architecture} \label{supsec_theory}
\setcounter{equation}{0} \setcounter{figure}{0} 
\setcounter{table}{0} 
We first introduce some concepts and notation. A function $f: \mathbb{R}^d \to \mathbb{R}$ is called $(\wp, C)$-smooth for \( \wp = k + \beta \) with $k \in \mathbb{N}_+$ and \( \beta \in (0,1] \) if for every \( \bdalpha = (\alpha_1, \dots, \alpha_d)^\top \) with \( |\alpha| = \sum_{i=1}^d \alpha_i \leq k \), the partial derivative \( D^\bdalpha f \) exists and satisfies the H\"older condition with exponent \( \beta \) and constant \( C \):
    \[
    |D^\bdalpha f(\bdx) - D^\bdalpha f(\bdy)| \leq C \|\bdx - \bdy\|^\beta \quad \forall \bdx, \bdy \in \mathbb{R}^d.
    \]
    
Under the basis function representation described in Sections 2.3 and 2.4, the SVFIM of order $d^\ast$ and level $0$ described in (2.2) can be rewritten as $\E \{Y(\bds) | \BX(\bds; \cdot), \BZ(\bds)\} = g\{ v_1(\bds), \ldots, v_{d^\ast}(\bds)\}$ where
\begin{eqnarray}\label{eqn:fmim_neuron_basis}
    && v_\ell(\bds) = \sum_{k=1}^K \sum_{m=1}^{M_k} \sum_{p=1}^P \kappa_{\ell kmp} \psi_p(\bds) \int_\mathcal{T}f_{ km}(t) X_{k}(\bds;t)\mathrm{d}t \nonumber \\
 && \hskip20mm + \sum_{j=1}^J \sum_{p=1}^P \vartheta_{\ell jp}  \psi_p(\bds) Z_j(\bds) + \sum_{h=1}^H \gamma_{\ell h} \phi_h(\bds), 
\end{eqnarray}
$\ell=1, \ldots, d^\ast$. With the basis function representation, the general SVFIM defined in (2.3) can be written as
$\E \{Y(\bds) | \BX(\bds; \cdot), \BZ(\bds)\} = \MFM\{\CX(\bds)\}$, where $\CX(\bds)=(\CX_1, \CX_2, \CX_3)^\top(\bds)$, with $\CX_1(\bds)$ $=$ $\{\psi_p(\bds) \int_\mathcal{T}f_{km}(t) X_{k}(\bds;t)\mathrm{d}t,$ $\ p\in [P],\ m\in [M_k],\ k\in [K]\}^\top$, $\CX_2(\bds)=\{ \psi_p(\bds) Z_j(\bds), \ j\in [J], \ p\in [P]\}^\top $ and $\CX_3(\bds)=\{ \phi_h(\bds), \ h\in [H]\}^\top $, and $\MFM(\cdot)$ belongs to the class of generalized hierarchical interaction models considered by \cite{BauerKohler2019aos}. 

Let $D$ be the dimension of $\CX$, and define a class of two-layer neural networks $\CF_{M^\ast, d^\ast, D, \tau}$ which is the set of all functions $f: \mathbb{R}^D \to \mathbb{R}$ of the form
\bse
    f(\bdx)= \sum_{i=1}^{M^\ast} w_i^{[2]} \sigma\bigg( \sum_{j=1}^{4 d^\ast} w_{ij}^{[1]} \sigma \bigg( \sum_{v=1}^D w_{ijv}^{[0]} x_v +b_{ij}^{[0]}\bigg)+ b_i^{[1]} \bigg) + b^{[2]}, 
\ese
with $\max_{i,j,v} \{ |w_{ijv}^{[0]}|, |w_{ij}^{[1]}|, |w_{i}^{[2]}|, |b_{ij}^{[0]}|, |b_{i}^{[1]}|, |b^{[2]}|\}< \tau $. Let $\wh \MFM(\CX) \in \CF_{M^\ast, d^\ast, D, \tau}$ be the neural network estimator of $\MFM(\CX)$ that minimizes the least square loss in a training set of sample size $n$. Set  $M^\ast \asymp n^{ d^\ast \over 2\wp +d^\ast} $ and $\tau \asymp n^c$ for some constant $c>0$. Suppose $\{ \CX(\bds)\}$ is a stationary random field and $\MFM(\CX)$ is a $(\wp, C)$-smooth hierarchical interaction model, then following \cite{BauerKohler2019aos}
\ben
    \E[ \wh \MFM\{\CX(\bds)\} -\MFM\{\CX(\bds)\}]^2 \le C \log(n)^3 n^{d^\ast \over 2\wp +d^\ast}.
\een
This result implies that when a low-rank structure such as the SVFIM holds, the proposed SVD-funNet does not suffer from the curse of dimensionality in the sense that the convergence rate depends on order $d^\ast$, which is much lower than the dimension of the input $D$.
\section{Additional Figures from Simulation Experiments}
\begin{figure}[t!]
\centering
    \includegraphics[width=5.5in]{ 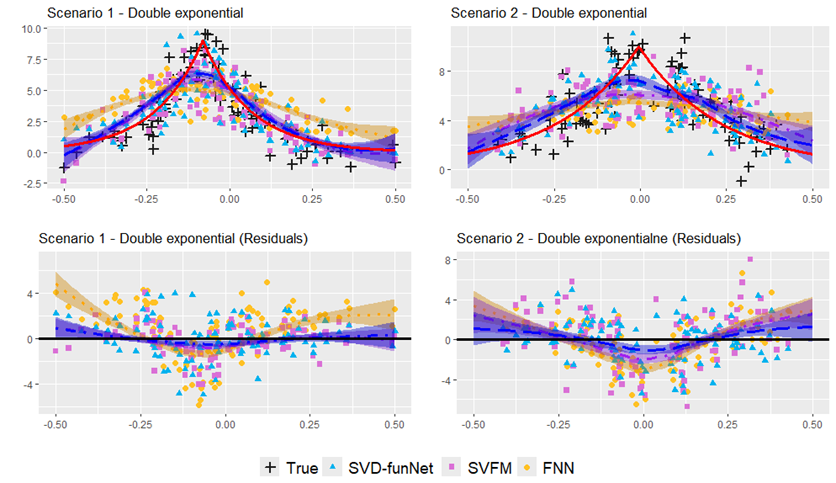}\\[1cm]
      \includegraphics[width=5.5in]{ 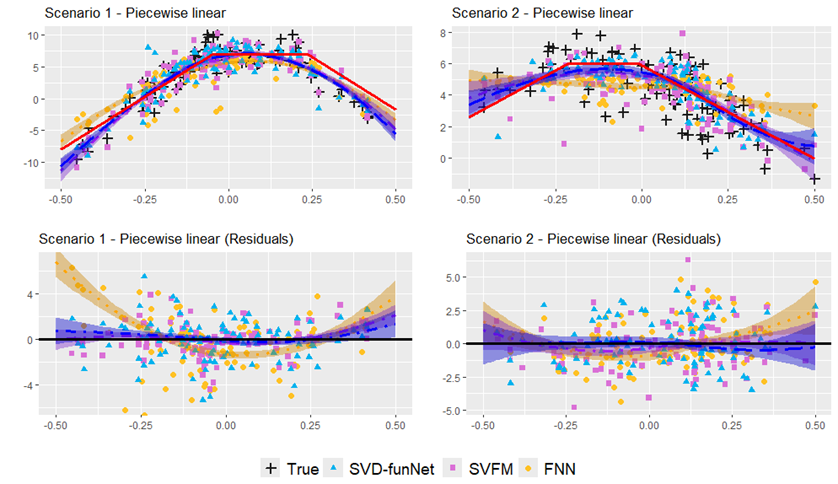}
   \caption{\small For each choice of $g(\cdot)$, top panel: Estimated $g$ function from SVD-funNet (blue), SVFM (purple), and FNN (yellow) under Scenarios 1 and 2, based on a randomly selected simulation run with SNR = 2.5. To aid visualization, 100 data points from the test set are randomly selected for display. The x-axis represents ${Z_k(\bds_l) \alpha(\bds_l) + \int_{\mathcal{T}} X_k(\bds_l;t) \beta(\bds_l;t) \textrm{d}t + \eta(\bds_l)}$ from the equation (3.1) of the manuscript, and the y-axis represents $\hat{Y}_k(\bds_l)$. For clarity, the data points are centered and scaled, and local regression curves (long dashed lines) with corresponding standard error bands are superimposed. The solid red line indicates the true sine function for $g(\cdot)$, and the black "+" symbols denote the true values of $Y_k(\bds_l)$.
  Bottom panel: Corresponding residual plots for the top panel, also superimposed with local regression curves. The solid black line represents the zero line on the y-axis. } \label{figure:sim_link_supp} 
\end{figure}

Figures~\ref{figure:sim_link_supp} illustrate the recovery of the double exponential and piecewise linear functions $g(\cdot)$, respectively, by each method under Scenarios 1 and 2. Consistent with the results shown in Figure 3 in Section 3.3, both SVFM and SVD-funNet successfully recover the nonlinear function $g(\cdot)$. Under Scenario 1, the residuals from SVD-funNet exhibit slightly greater variance compared to those from SVFM. However, under Scenario 2, the residuals from SVD-funNet are more centered around zero and display slightly lower variability than those from SVFM.

\section{Additional Figures from the Crop Yield Prediction Application}
\begin{figure}[h!]
  \centering
  \includegraphics[width=6in]{ 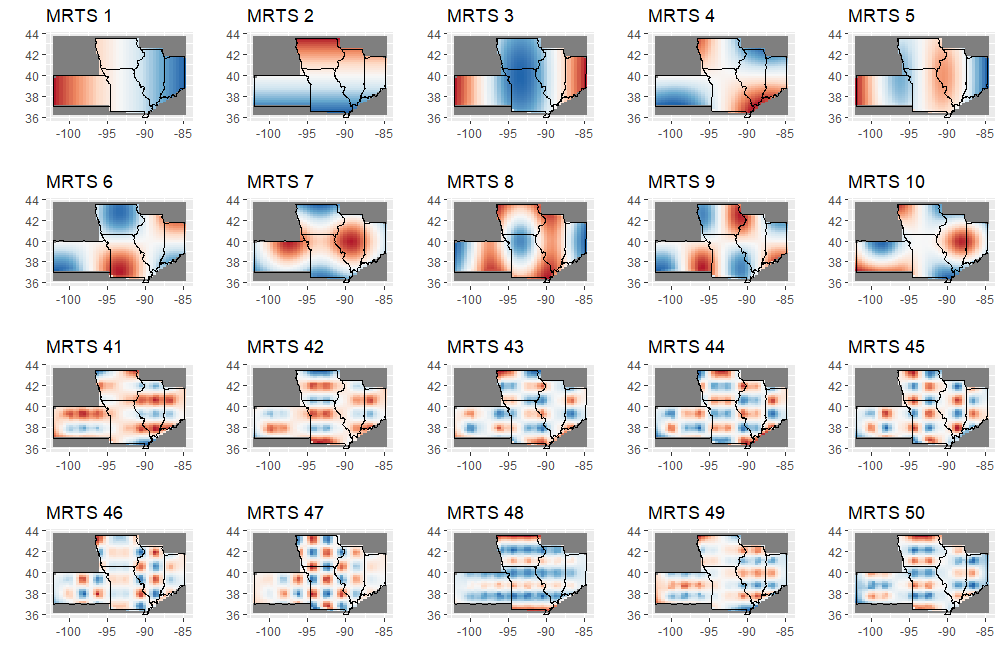}
  \caption{Illustration of the first 10 MRTS basis functions, labeled as MRTS 1 to 10, displaying global spatial structure, and 41st to 50th MRST basis functions, labeled as MRTS 41 to 50. Evaluated values are normalized, with dark red indicating larger positive values, dark blue indicating smaller negative values, and white representing zero.}
  \label{fig:MRTS}
\end{figure}

\begin{figure}[t!]
  \centering
  \includegraphics[width=5.5in]{ 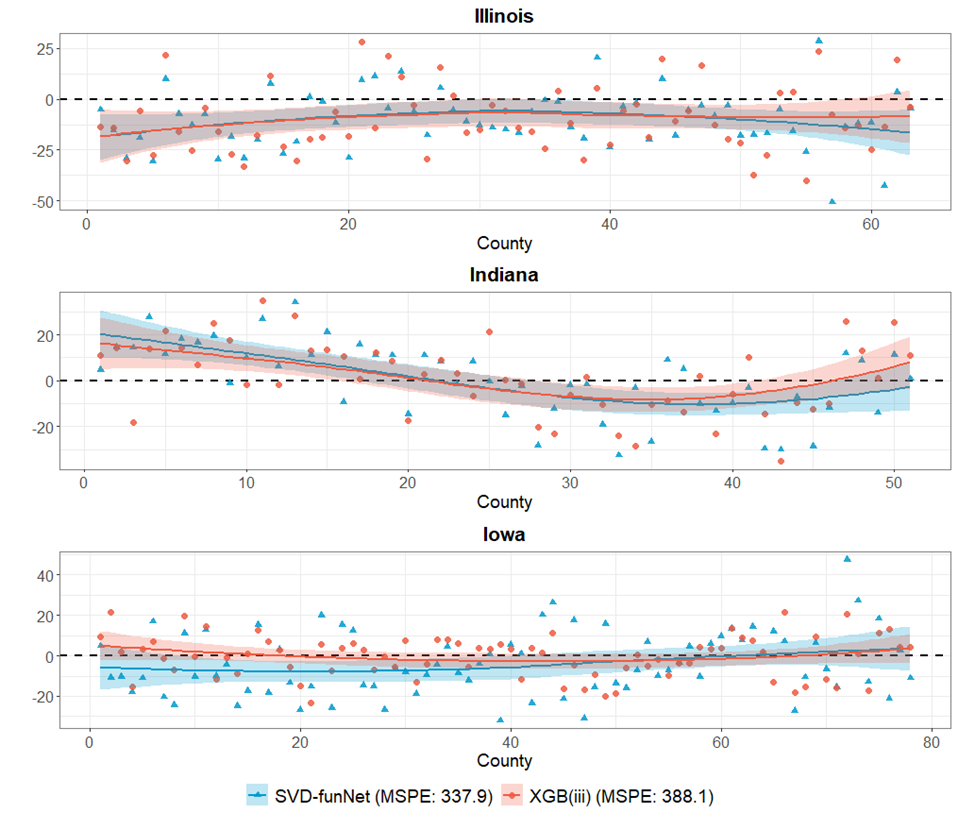}
  \caption{\black{Prediction residuals across counties in Illinois, Indiana, and Iowa from the model to predict 2018 corn yields using 1999-2017 data from the proposed SVD-funNet (blue) and the XGB(iii) method enhanced with spatially varying weights on FPC scores and spatial random effects (red). The corresponding local regression curves are superimposed, with shaded regions indicating their associated standard errors.}}
  \label{figure:prediction_three_states}
  \vspace{1.5cm}
\end{figure}

\begin{figure}[t!]
  \centering
  \includegraphics[width=6in]{ 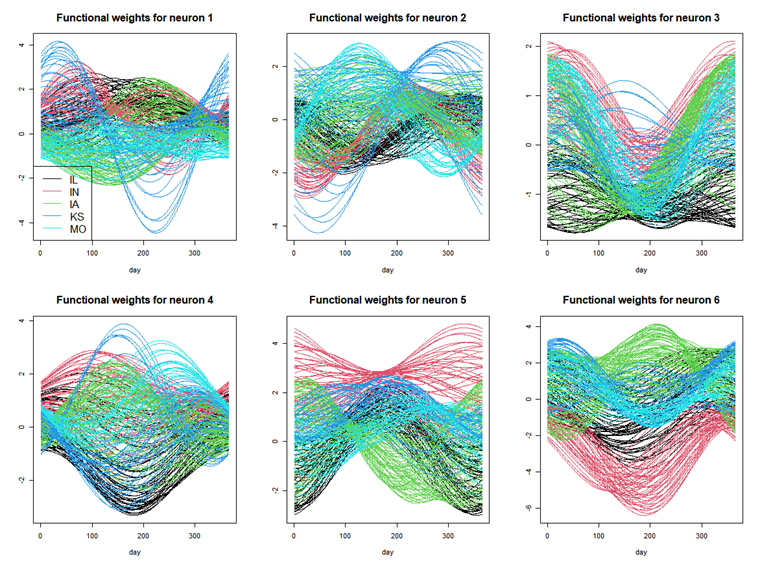}
  \caption{Illustration of functional neural network weights for maximum temperature trajectories from the first six neurons in the first hidden layers. Functional weights for counties in Illinois, Indiana, Iowa, Kansas, and Missouri are highlighted in black, red, green, blue, and turquoise, respectively.  }
  \label{figure:fweight}
\end{figure}

Figure \ref{fig:MRTS} illustrates the first 10 MRTS basis functions, which capture global variations, alongside the 41st to 50th MRTS basis functions, which capture local variations, based on 40 equally spaced inner knots selected from the spatial domain in the real data application. Table \ref{tab:statewise_mspe} reports the statewise MSPE and weighted MSPE for SVD-funNet and XGB predictions of 2018 corn yield, based on training data from 1999 to 2017 across five Midwestern states. 
\black{Figure~\ref{figure:prediction_three_states} illustrates county-level prediction residuals from the proposed SVD-funNet and the XGB model, overlaid with local regression curves and shaded bands representing their standard errors, for Illinois, Indiana, and Iowa. The models are trained using data from 1999 to 2017 to predict crop production in 2018. Consistent with the results in Table~\ref{tab:statewise_mspe}, the XGB model performs comparably to SVD-funNet in Illinois and Indiana, while slightly outperforming SVD-funNet in Iowa, with residuals remaining closer to zero.}
Lastly, Figure \ref{figure:fweight} illustrates functional weights for maximum temperature trajectories from the first six neurons, estimated from a model with four hidden layers with 32 neurons for each under the ‘sigmoid’ activation function. These weights highlight the degree of spatial variability in the model parameters. 

\newpage
\begin{table}[h!]
\centering
\caption{MSPE and Weighted MSPE for SVD-funNet and XGB predictions for 2018 corn yield across five Midwest states.} \label{tab:statewise_mspe}
\begin{tabular}{l|cc|cc}
\toprule
\multirow{2}{*}{State} & \multicolumn{2}{c|}{MSPE} & \multicolumn{2}{c}{Weighted MSPE} \\
\cmidrule(lr){2-3} \cmidrule(lr){4-5}
 & SVD-funNet & XGB & SVD-funNet & XGB \\
\midrule
Illinois(IL) & 306.4 & 360.2 & 283.9 & 350.0 \\
Kansas(KS) & 686.1 & 1153.6 & 636.8 & 991.0 \\
Missouri(MO) & 255.7 & 336.2 & 211.3 & 314.9 \\
Iowa(IA) & 260.9 & 107.4 & 242.0 & 105.9 \\
Indiana(IN) & 257.6 & 256.0 & 232.8 & 231.5 \\
\bottomrule
\end{tabular}
\end{table}

\newpage

\newpage

\newpage

\end{document}